# Field-free deterministic switching of a perpendicularly polarized magnet using unconventional spin-orbit torques in WTe$_2$

I-Hsuan Kao[1,2], Ryan Muzzio[1,2], Hantao Zhang[3], Menglin Zhu[4], Jacob Gobbo[1], Daniel Weber[5,6], Rahul Rao[7], Jiahan Li[8], James H. Edgar[8], Joshua E. Goldberger[5], Jiaqiang Yan[9,10], David G. Mandrus[9,10], Jinwoo Hwang[4], Ran Cheng[3,11], Jyoti Katoch[1], and Simranjeet Singh[1,*]

[1]*Department of Physics, Carnegie Mellon University, Pittsburgh, PA, 15213, USA*

[3]*Department of Electrical and Computer Engineering, University of California, Riverside, CA 92521, USA*

[4]*Department of Materials Science and Engineering, The Ohio State University, Columbus, OH 43210, USA*

[5]*Department of Chemistry, The Ohio State University, Columbus, OH 43210, USA*

[6]*Battery and Electrochemistry Laboratory (BELLA), Institute of Nanotechnology, Karlsruhe Institute of Technology (KIT), Hermann-von-Helmholtz Platz 1, 76344 Eggenstein-Leopoldshafen, Germany*

[7]*Materials and Manufacturing Directorate, Air Force Research Laboratory, Wright-Patterson Air Force Base, Dayton, OH, 45433, USA*

[8]*Tim Taylor Department of Chemical Engineering, Kansas State University, Manhattan, Kansas 66506, USA*

[9]*Materials Science and Technology Division, Oak Ridge National Laboratory, Oak Ridge, Tennessee 37831, USA*

[10]*Department of Materials Science and Engineering, The University of Tennessee, Knoxville, TN 37996, USA*

[11]*Department of Physics and Astronomy, University of California, Riverside, CA 92521, USA*

[2] *These authors contributed equally to this work: I. Kao, R. Muzzio*

[*]*Email: simranjs@andrew.cmu.edu*

**Spin-orbit torque (SOT) driven deterministic control of the magnetization state of a magnet with perpendicular magnetic anisotropy (PMA) is key to next generation spintronic applications including non-volatile, ultrafast, and energy efficient data storage devices[1-3]. But, field-free deterministic switching of perpendicular magnetization remains a challenge because it requires an out-of-plane anti-damping torque, which is not allowed in conventional spin source materials such as heavy metals (HM)[1-3] and topological insulators[4] due to the system's symmetry. The exploitation of low-crystal symmetries in emergent quantum materials offers a unique approach to achieve SOTs with unconventional forms[5,6]. Here, we report the first experimental realization of field-free deterministic magnetic switching of a perpendicularly polarized van der Waals (vdW) magnet employing an out-of-plane anti-damping SOT generated in layered $WTe_2$ which is a low-crystal symmetry quantum material. The numerical simulations confirm that out-of-plane antidamping torque in $WTe_2$ is responsible for the observed magnetization switching in the perpendicular direction.**

SOT has emerged as an efficient means of manipulating the magnetization state of a ferromagnetic (FM) material[1,2]. The potential technological applications based on SOT driven magnetization manipulation includes energy efficient non-volatile magnetic memories[7] and spin-torque oscillators[8,9]. In SOT induced magnetic switching, a charge current density $\vec{j}_c$ (*x* direction) flowing in HM/FM bilayers results in spin current flowing in the out-of-plane direction (*z* direction) via spin galvanic effects[1-3], which in turn exerts a torque on the magnetization of a nearby magnetic layer. The torque has an antidamping component $\vec{\tau}_{AD} \propto \vec{m} \times \vec{m} \times \vec{p}$ and a field-like component $\vec{\tau}_{FL} \propto \vec{m} \times \vec{p}$ where $\vec{m}$ is the magnetization, and $\vec{p}$ is the spin polarization. But by symmetry, the most efficient type of torque, the anti-damping torque, has the form of $\vec{\tau}_{IP}^{AD} \propto \vec{m} \times \vec{m} \times \hat{y}$ in HM/FM heterostructures and can only deterministically switch the magnetization of a magnet that has an in-plane magnetic anisotropy[3,10].

However, for memory applications, magnets with PMA are highly desired because it allows for ultra-compact packing and thermally stable nanometer sized magnetic bits[7]. In conventional HM/FM systems, a small external magnetic field is applied along the direction of the charge current to break the in-plane symmetry of the system allowing for deterministic switching of the magnetization state of a PMA magnet[1,3]. Previously, a structural asymmetry[11,12], tilted anisotropy of the nanomagnet[13], an in-plane magnetized layer[14-16], and an in-plane effective magnetic field in FM/ferroelectric structures[17], anomalous Hall effect and planar Hall effect in FM/HM/FM tri-layers[18], and interplay of SOT and spin-transfer torque[19], have been explored to achieve a field-free deterministic switching of PMA magnets using SOTs. A novel approach for this is to explore the utility of emergent quantum systems as a spin source material wherein SOTs can be controlled by crystal symmetries. Recently, the transition metal dichalcogenides with low-crystal symmetry, such as $WTe_2$, have been shown to exhibit an out-of-plane antidamping torque, $\vec{\tau}_{OP}^{AD} \propto \vec{m} \times \vec{m} \times \hat{z}$, when current is applied along the low symmetry axis of the $WTe_2$/FM bilayer system[5,6]. But, so far, there has been no experimental study establishing that the unconventional out-of-plane antidamping SOT in layered materials with low-crystal symmetry is strong enough to enable the deterministic switching of a PMA magnet.

Here, we experimentally realize field-free deterministic magnetic switching of a perpendicularly polarized magnet using SOTs in a low-symmetry system. For this, we employ a SOT device platform built out of vdW based layered quantum materials and choose $WTe_2$ as a spin-source material for generating SOTs[5,20-23]. In addition to the generation of an out-of-plane antidamping torque, WTe$_2$ also exhibits properties that are highly relevant for a large charge to spin conversion efficacy: strong spin-orbit coupling, non-trivial band dispersion, topologically protected spin polarized bulk and surface states, pronounced Edelstein effect, and an intrinsic spin Hall effect[20-23]. WTe$_2$ is a low-symmetry system, whose *ab*-plane is schematically depicted in Fig. 1a (*c*-axis is perpendicular to the *ab*-plane). The surface of WTe$_2$ has a mirror symmetry with respect to the *bc*-plane but not about the *ac*-plane. Thus, the system is asymmetric relative to a 180° rotation about the *c*-axis. For the PMA magnet, we utilize Fe$_{2.78}$GeTe$_2$ (FGT) which is a layered vdW FM material[24,25]. The SOT devices are prepared by mechanical dry transfer technique to assemble WTe$_2$/FGT bilayers and standard device fabrication techniques (Methods and Supplementary Note 1).

In total, we have measured three devices and here we present data from device A. The optical micrograph of the device is shown in Fig. 1b along with a side-view schematic of the device (lower panel). The crystallographic *a*-axis and *b*-axis are labeled in Fig. 1b, and are confirmed by polarized Raman spectroscopy[26]. Spectra are collected by rotating the polarization of the incident laser for different angles relative to the *a*-axis of WTe$_2$ and the integrated intensities under each peak are calculated for the contour plot shown in Fig. 1c. A polar plot of the A$_g$ Raman peak at 212 cm$^{-1}$ (Fig. 1d, corresponding to the dashed line in Fig. 1c) exhibits minimum intensity when the excitation laser polarization is along the straight edge of the WTe$_2$ flake (Fig. 1b). This is consistent with previous reports[22,26] and clearly distinguishes the *a*-axis of WTe$_2$. We also carried out atomic structure characterization of device A using scanning transmission electron microscopy (STEM). Fig. 1e (bottom panel) shows the STEM cross-sectional image confirming that the lateral direction is the *a*-axis WTe$_2$, consistent with the orientation determined by the Raman spectra. The top-panel of Fig. 1e schematically shows the atomic arrangement of WTe$_2$ along the *a*-axis as a comparison to the STEM image.

Previously, the presence of a strong out-of-plane antidamping torque in a WTe$_2$/Permalloy heterostructure was probed by spin torque ferromagnetic resonance[5]. Unlike an out-of-plane field-like torque, the out-of-plane antidamping torque is independent of the reversal of magnetization, reverses with current direction, and can efficiently switch perpendicular magnetizations. The out-of-plane antidamping torque is not allowed in conventional spin source systems, such as platinum and topological insulators, due to in-plane 2-fold symmetry. However, in WTe$_2$ this symmetry is broken around one crystal axis, allowing for an out-of-plane antidamping torque. Specifically, WTe$_2$ has a broken mirror symmetry about the *ac*-plane (Fig. 1a), allowing for this torque to exist only when the current is driven along the *a*-axis of WTe$_2$ (Fig. 2a, upper panel). Along the *b*-axis of WTe$_2$, the mirror symmetry is preserved and the out-of-plane antidamping torque cannot exist (lower panel of Fig. 2a): when a charge current passes along the *b*-axis and a *bc*-plane mirror operation ($\tilde{R}_{bc}$) is performed, the current will not switch direction but the torque will still change signs in the out-of-plane direction. A current-induced torque must change sign with current otherwise the torque will have zero magnitude, therefore the out-of-plane antidamping torque vanishes when current is along the *b*-axis.

A typical anomalous Hall effect (AHE) hysteresis loop of the device A is shown in Fig. 2b (ref. [24,27]). We determine the effective magnetic anisotropic field $\mu_0 H_k \sim 4.12$ T by measuring the AHE loop with the field applied along the hard axis, indicating a strong PMA (Supplementary Note 4 and Note 6). Next, we demonstrate field-free perpendicular magnetization switching induced by SOTs when a DC current pulse, $I_p$, is applied (Methods). For the case when current is applied along the *a*-axis ($\vec{I} \parallel \hat{a}$), we observe clear deterministic switching (Fig. 2c) which is attributed to the out-of-plane spin polarization ($p_z$) or an out-of-plane antidamping SOT. The measurements (current sweeps in Fig. 2c) show that the terminal state is determined by the polarity of $\vec{I}$ and the critical current to switch the magnetization is $I_c = 5.63 \pm 0.13$ mA (corresponding to a critical current density $J_c = 6.64 \times 10^{10}$ A/m$^2$). On the other hand, the behavior is completely different when the current is applied along the *b*-axis; no deterministic switching is observed in the absence of an external in-plane magnetic field (Fig. 2d). In the high current regime, $m_z$ approaches zero with increasing $I_p$, reflecting multi-domain formation due to Joule heating. This case is similar to SOT switching in conventional HM/FM bilayer systems: the terminal state is close to the demagnetized state $m_z \approx 0$, indicating that the in-plane effective field, generated by either the Oersted effect or spin galvanic effects, drives the magnetic moment of FGT to the device plane. Without the presence of a $p_z$ (or a symmetry breaking in-plane magnetic field), multi-domains are formed after the current is turned off. Each domain is then left randomly oriented which leads to the demagnetization of FGT. Furthermore, using a train of positive and negative current pulses applied along the *a*-axis of WTe$_2$ (top panel of Fig. 2e) we deterministically switch the magnetic state of the FGT from up to down and vice-versa (bottom panel of Fig. 2e).

To distinguish the role of the in-plane antidamping torque $\vec{\tau}_{IP}^{AD}$ and the out-of-plane antidamping torque $\vec{\tau}_{OP}^{AD}$, we perform numerical simulations of magnetization switching for a mono-domain FGT. While the actual switching processes are likely to involve domain-wall motions, a monodomain simulation can capture the essential physics of spin torques if we use the coercive field (Supplementary Information Fig. S6) as the effective anisotropy. We first investigate the case of $\vec{I} \parallel \hat{b}$. As shown in Fig. 2f, by varying the current $I_e$ and the ratio of $\vec{\tau}_{IP}^{AD}/I_e$, we find that the initial state ($m_z = -1$) can be switched to an in-plane direction ($m_z = 0$) that are energetically metastable. The threshold decreases with an increasing $\vec{\tau}_{IP}^{AD}/I_e$ which forms the phase boundary. However, if the driving current is turned off, a metastable state will be equally probable to relax to $m_z = -1$ and $m_z = 1$ The mono-domain will easily break up into multiple domains with an average $<m_z> = 0$. This is consistent with Fig. 2d, justifying that an in-plane torque $\vec{\tau}_{IP}^{AD}$ alone is unable to switch the magnetization from $m_z = -1$ to $m_z = 1$. By comparing Fig. 2f with Fig. 2d, we can determine the value of $\vec{\tau}_{IP}^{AD}/I_e$. Then, with a fixed ratio of $\vec{\tau}_{OP}^{AD}/I_e$, we simulate the switching process for $\vec{I} \parallel \hat{a}$ by varying $I_e$ and $\vec{\tau}_{OP}^{AD}/I_e$. As shown in Fig. 2g, we find three different phases characterized by the final state of $m_z$: the unswitched phase ($m_z = -1$), the switched phase ($m_z = 1$), and the in-plane metastable phase ($m_z = 0$). The three phases intersect at a tricritical point around $I_c = 5.9$ mA and $\vec{\tau}_{OP}^{AD}/I_e = 6.2 \times 10^{-5}$ T/mA. Above the tricritical point, the phase boundary separating the unswitched and the switched phases continuously shifts towards smaller $I_e$, indicating that the out-of-plane antidamping torque $\vec{\tau}_{OP}^{AD}$ is indeed responsible for the magnetization switching in the perpendicular direction. This result not only explains the observation in Fig. 2c but also determines the relative strength of $\vec{\tau}_{OP}^{AD}$ over $\vec{\tau}_{IP}^{AD}$ to be around $10^{-2}$ for $\vec{I} \parallel \hat{a}$.

To examine the existence of a $p_z$, we also perform AHE loop shift measurements. In the presence of a $p_z$, an out-of-plane antidamping SOT can be generated and abruptly shift the AHE hysteresis loop once the current passes a threshold value such that the intrinsic damping is compensated[1,18]. For $\vec{I} \parallel \hat{a}$ (Fig. 3a) and when $I_p$ is positive (negative), the AHE hysteresis loop shifts to the negative (positive) side. However, for the same conditions but with $I_p$ applied along the *b*-axis (Fig. 3b), no significant loop shift is observed. The measurements are performed at different current magnitudes for $\vec{I} \parallel \hat{a}$ and $\vec{I} \parallel \hat{b}$ (Fig. 3c and 3d). Strikingly, we observe two threshold currents and the signature of magnetic excitations due to spin current induced steady precessional states[28]. Around the first current threshold ($I_{th1} \sim 3.5$ mA), there is a peak in $H_c$ which decreases dramatically with increasing $|I_p|$. The peak of $H_c$ is accompanied by a sudden increase in $|H_{sh}|$ pointing to the onset of magnetic excitation[28]. The abrupt drop of $H_c$ is the sign of the system entering a current-assisted switching regime, which was seen in a similar measurement setup in a conventional SOT PMA system[1]. Joule heating can also contribute to the drop of $H_c$ (Supplementary Note 5). When $|I_p|$ passes the second current threshold ($I_{th2} \sim 5.75$ mA), we observe the onset of a current-induced switching regime, where the decrease in $H_c$ levels off and, particularly for $\vec{I} \parallel \hat{a}$, the deterministic switching is possible because $H_c^{\pm}$ both shift to one side of the zero field line. Note that $I_{th2}$ is very close to $I_c$ which is determined by current-induced SOT switching at zero field (Fig. 2c). In contrast, for $\vec{I} \parallel \hat{b}$, $|H_{sh}(I_p)|$ shows very weak $I_p$ dependence and is close to zero at all current values which is similar to what has been reported for conventional HM/FM systems[1]. These results clearly show the presence of a $p_z$ when $\vec{I} \parallel \hat{a}$, originating from the broken mirror symmetry about the *ac*-plane. The abrupt shift, instead of a linear shift, in $H_{sh}$ also indicates that no significant out-of-plane field-like torque is present in the system.

To conclude, we have experimentally demonstrated that an out-of-plane antidamping SOT in WTe$_2$ enables a field free perpendicular magnetization switching of PMA magnet. The presence of an out-of-plane antidamping torque and the potential of a higher charge to spin conversion efficiency in WTe$_2$ makes TMDs with lower symmetry crystal structure an appealing spin source material for future SOT related magnetic technologies.

## Methods

**Device Fabrication.** WTe$_2$, FGT, and hexagonal boron nitride (hBN) crystal flakes were prepared by previously published procedures[25,29,30]. WTe$_2$, hBN, and FGT were mechanically exfoliated on separate Si/SiO$_2$(300nm) substrates inside an Argon glovebox environment. The flakes were then optically searched and selected using a fully automated microscope inside the glovebox. On a separate substrate, electrodes were defined using standard electron beam lithography with MMA/PMMA bilayer resist and electron beam deposition was used for Pt electrodes. The heterostructure was made using a custom-built transfer tool inside the glovebox using a polydimethylsiloxane (PDMS) stamp and thin film of polycarbonate (PC). The final device consists, from top to bottom, of hBN/FGT/WTe$_2$/Pt with Cr(5nm)/Au(110nm) bond pads.

**Spin orbit torque measurements.** The electrical measurements were performed at variable temperatures in high vacuum (pressure $< 10^{-5}\ mTorr$) conditions. An electromagnet was rotated such that the magnetic field can be applied in both in- and out-of-plane directions of the device. Keithley 6221 current source and a Keithley 2182A nanovoltmeter are used for AHE hysteresis loop and current pulse-induced SOT switching experiments. The current pulse used is a square current pulse with varying magnitude and a width of $100\ \mu s$ (unless stated otherwise). The transverse resistance, $R_{xy}$, is measured with a smaller magnitude current ($50\ \mu A$) in delta mode (unless stated otherwise) to determine the magnetization state of FGT. We define the normalized perpendicular magnetization $m_z = \frac{M_z}{M_s} = \frac{R_{xy}}{R_{AHE}}$, where $M_z$ is the perpendicular magnetization, and $M_s$ is the saturation magnetization. In the SOT measurements, the initial magnetic state is prepared in $m_z = \pm 1$ (using an external magnetic field) and $I_p$ is swept from $0\ mA$ to $\pm 10\ mA$ in steps of $250\ \mu A$. We define the critical switching current $I_c = \frac{I_c^+ + I_c^-}{2}$, where $I_c^+$ ($I_c^-$) is the current pulse magnitude that drives the magnetization state across $m_z = 0$ at positive (negative) pulse sides. For SOT switching with pulse trains, a series of read current pulses of 500 $\mu A$ were applied to read the magnetization state before and after the write current pulse was applied. AHE loop shift measurements are performed by measuring $m_z(H_z)$, where at each perpendicular field $H_z$ a current pulse $I_p$ ($500\ \mu s$ long) is applied and $m_z$ is measured simultaneously in Pulse Delta mode. We define the coercive field by $H_c = (H_c^+ - H_c^-)/2$ and the shift of the loop by $H_{sh} = [H_c^+ + H_c^-]/2$, where $H_c^+$ ($H_c^-$) is the positive (negative) field for which the magnetization reverses.

**Numerical simulations.** The effective single-domain dynamics is simulated by solving the Landau–Lifshitz–Gilbert equation in the presence of in-plane and out-of-plane antidamping torques as

$$\frac{d\mathbf{m}}{dt} = -\gamma H_A \mathbf{m} \times m_z \hat{\mathbf{z}} + c_{oe} I_e \hat{\mathbf{y}} \times \mathbf{m} + \mathbf{m} \times \left[ \left( c_{IP}^{AD} I_e \hat{\mathbf{y}} + c_{OP}^{AD} I_e \hat{\mathbf{z}} \right) \times \mathbf{m} \right] + c_{OP}^{FL} I_e \hat{\mathbf{z}} \times \mathbf{m} + \alpha \mathbf{m} \times \frac{d\mathbf{m}}{dt}$$

where $\mathbf{m}$ is the unit vector of magnetic moment of FGT, $H_A$ is the easy-axis anisotropy, $I_e$ is the current, $c_{oe}$, $c_{IP}^{AD}$, $c_{OP}^{AD}$ and $c_{OP}^{FL}$ are the amplitudes of the Oersted field, the in-plane antidamping torque, the out-of-plane antidamping torque and the out-of-plane field-like torque with respect to $I_e$, respectively, and $\alpha$ is the Gilbert damping parameter. Based on the materials parameters and the device geometry, we set $H_A = 0.1T$ (using the coercive field as the effective value), $c_{oe} = 2 \times 10^{-4}\ T/mA$, $\alpha = 0.005$. In the simulation, we use $c_{OP}^{FL} = 0.5\ c_{OP}^{AD}$, but we find that $c_{OP}^{FL}$ has negligible impact on the switching process; the result does not show visible differences if $c_{OP}^{FL} = 0$ or $c_{OP}^{FL} = 4\, c_{OP}^{AD}$. The current $I_e$ is applied along $\hat{y}$ direction, so the in-plane spin accumulation and the Oersted field are along $\hat{x}$ direction. The initial state of $\boldsymbol{m}$ is set to be along $-\hat{z}$ direction, and the final state is obtained when the system reaches equilibrium.

## References

1. Miron, I. M. *et al.* Perpendicular switching of a single ferromagnetic layer induced by in-plane current injection. *Nature* **476**, 189-193, (2011).
2. Liu, L. *et al.* Spin-Torque Switching with the Giant Spin Hall Effect of Tantalum. *Science* **336**, 555-558, (2012).
3. Liu, L. *et al.* Current-Induced Switching of Perpendicularly Magnetized Magnetic Layers Using Spin Torque from the Spin Hall Effect. *Physical Review Letters* **109**, 096602, (2012).
4. Mellnik, A. R. *et al.* Spin-transfer torque generated by a topological insulator. *Nature* **511**, 449-451, (2014).
5. MacNeill, D. *et al.* Control of spin–orbit torques through crystal symmetry in WTe$_2$/ferromagnet bilayers. *Nature Physics* **13**, 300-305, (2017).
6. Xue, F. *et al.* Unconventional spin-orbit torque in transition metal dichalcogenide--ferromagnet bilayers from first-principles calculations. *Physical Review B* **102**, 014401, (2020).
7. Brataas, A. *et al.* Current-induced torques in magnetic materials. *Nature materials* **11**, 372-381, (2012).
8. Liu, R. H. *et al.* Spectral Characteristics of the Microwave Emission by the Spin Hall Nano-Oscillator. *Physical Review Letters* **110**, 147601, (2013).
9. Cheng, R. *et al.* Terahertz Antiferromagnetic Spin Hall Nano-Oscillator. *Physical Review Letters* **116**, 207603, (2016).
10. Garello, K. *et al.* Symmetry and magnitude of spin–orbit torques in ferromagnetic heterostructures. *Nature Nanotechnology* **8**, 587-593, (2013).
11. Yu, G. *et al.* Switching of perpendicular magnetization by spin–orbit torques in the absence of external magnetic fields. *Nature Nanotechnology* **9**, 548-554, (2014).
12. Chen, S. *et al.* Free Field Electric Switching of Perpendicularly Magnetized Thin Film by Spin Current Gradient. *ACS Applied Materials & Interfaces* **11**, 30446-30452, (2019).
13. You, L. *et al.* Switching of perpendicularly polarized nanomagnets with spin orbit torque without an external magnetic field by engineering a tilted anisotropy. *Proceedings of the National Academy of Sciences* **112**, 10310-10315, (2015).
14. Lau, Y.-C. *et al.* Spin–orbit torque switching without an external field using interlayer exchange coupling. *Nature Nanotechnology* **11**, 758-762, (2016).
15. Fukami, S. *et al.* Magnetization switching by spin–orbit torque in an antiferromagnet–ferromagnet bilayer system. *Nature materials* **15**, 535-541, (2016).
16. Oh, Y.-W. *et al.* Field-free switching of perpendicular magnetization through spin–orbit torque in antiferromagnet/ferromagnet/oxide structures. *Nature Nanotechnology* **11**, 878-884, (2016).
17. Cai, K. *et al.* Electric field control of deterministic current-induced magnetization switching in a hybrid ferromagnetic/ferroelectric structure. *Nature materials* **16**, 712-716, (2017).

18. Baek, S.-h. C. *et al.* Spin currents and spin–orbit torques in ferromagnetic trilayers. *Nature materials* **17**, 509-513, (2018).
19. Wang, M. *et al.* Field-free switching of a perpendicular magnetic tunnel junction through the interplay of spin–orbit and spin-transfer torques. *Nature Electronics* **1**, 582-588, (2018).
20. Feng, B. *et al.* Spin texture in type-II Weyl semimetal $WTe_2$. *Physical Review B* **94**, 195134, (2016).
21. Johansson, A. *et al.* Edelstein effect in Weyl semimetals. *Physical Review B* **97**, 085417, (2018).
22. Shi, S. *et al.* All-electric magnetization switching and Dzyaloshinskii–Moriya interaction in $WTe_2$/ferromagnet heterostructures. *Nature Nanotechnology* **14**, 945-949, (2019).
23. Zhao, B. *et al.* Unconventional Charge–Spin Conversion in Weyl-Semimetal $WTe_2$. *Advanced Materials* **32**, 2000818, (2020).
24. Fei, Z. *et al.* Two-dimensional itinerant ferromagnetism in atomically thin $Fe_3GeTe_2$. *Nature materials* **17**, 778-782, (2018).
25. Weber, D. *et al.* Decomposition-Induced Room-Temperature Magnetism of the Na-Intercalated Layered Ferromagnet $Fe_{3-x}GeTe_2$. *Nano Letters* **19**, 5031-5035, (2019).
26. Song, Q. *et al.* The polarization-dependent anisotropic Raman response of few-layer and bulk $WTe_2$ under different excitation wavelengths. *RSC Advances* **6**, 103830-103837, (2016).
27. Zheng, G. *et al.* Gate-Tuned Interlayer Coupling in van der Waals Ferromagnet $Fe_3GeTe_2$ Nanoflakes. *Physical Review Letters* **125**, 047202, (2020).
28. Lee, K. J. *et al.* Spin transfer effect in spin-valve pillars for current-perpendicular-to-plane magnetoresistive heads (invited). *Journal of Applied Physics* **95**, 7423-7428, (2004).
29. Zhao, Y. *et al.* Anisotropic magnetotransport and exotic longitudinal linear magnetoresistance in $WTe_2$ crystals. *Physical Review B* **92**, 041104, (2015).
30. Liu, S. *et al.* Single Crystal Growth of Millimeter-Sized Monoisotopic Hexagonal Boron Nitride. *Chemistry of Materials* **30**, 6222-6225, (2018).

## Acknowledgments

Funding for this research was provided primarily by the Center for Emergent Materials at The Ohio State University, a National Science Foundation (NSF) MRSEC through Award No. DMR-2011876. R.M. acknowledges the NSF support through AGEP-GRS supplement to award DMR-1809145. H.Z. and R.C. are supported by the Air Force Office of Scientific Research under grant FA9550-19-1-0307. J.Q.Y. acknowledges support from the U.S. Department of Energy, Office of Science, Basic Energy Sciences, Materials Sciences and Engineering Division. D.G.M. acknowledges support from the Gordon and Betty Moore Foundation's EPiQS Initiative through Grant No. GBMF9069. J.E.G. acknowledges the financial support from the Center of Emergent Materials, an NSF MRSEC, under the Grant No. DMR-2011876. D.W. gratefully acknowledges the financial support by the German Science Foundation DFG Research Fellowship (WE6480/1). Support for hBN crystal growth from the Office of Naval Research from award N00014-20-1-2427 is appreciated. Electron microscopy was performed at the Center for Electron Microscopy and Analysis at The Ohio State University.

## Author Contributions

S.S. and J.K. designed the experiments and supervised the research. I.K., R.M., and J.G. prepared the devices and performed the experiments. H.Z. and R.C. performed the numerical simulations. M.Z and J.H. performed the STEM measurements. R.R. carried out polarized Raman measurements. Y.J. and D.G.M. provided the bulk crystals of $WTe_2$. D.W. and J.E.G. provided the bulk FGT crystals. J.L and J. H. E. provided the bulk hBN crystals. All authors contributed to write the manuscript.

## Competing interests

The authors declare no competing interests.

# Figure Captions

**Fig. 1. Crystal structure, polarized Raman, and electron microscopy of the WTe2/ FGT device. a,** The crystal structure of WTe2 with *a*-axis and *b*-axis labeled. The crystal is invariant (noninvariant) upon an *bc* (*ac*) mirror operation. **b,** An optical image of the device A. The electrodes are set up to allow current pulses along the *a*- and b-axis of WTe2 separately. Lower panel: The side view of the device. **c,** The angle dependent polarized Raman spectra of WTe2 of the device A, obtained by rotating the laser polarization with respect to the long edge(*a*-axis) of WTe2 flake. **d,** The angle dependent Raman peak intensities of the Raman peak at 212 cm$^{-1}$, corresponds to dashed line in **(c)**. **e,** The top panel shows the atomic arrangement in WTe2 along the *a*-axis. Bottom panel: Cross-sectional STEM image of device A viewed along the *b*-axis orientation of WTe2, which confirms that the lateral direction is the *a*-axis of WTe2.

**Fig. 2. The spin-orbit torque induced switching in WTe2/ FGT bilayers. a,** Schematic to show how different spin-orbit torques change upon a bc-plane mirror operation (denoted by $\tilde{R}_{bc}$) depending on whether a charge current is applied along the a-axis or b-axis. **b,** the AHE hysteresis loop as a function of $H_z$ at $170\ K$. **c,d,** Current induced spin-orbit torque switching with magnetized state initialized in positive (negative) state and pulse sweeping from zero to large positive (negative) pulses while $\vec{I} \parallel \hat{a}$ **(c)** and $\vec{I} \parallel \hat{b}$**(d)**. **e,** The deterministic switching by a series of $500\ \mu s$ current pulses applied along the a-axis. The magnitude of the write pulse is 9 mA, corresponds to $J = 1.1 \times 10^{11} A/m^2$. **f,g,** Simulations for FGT single domain switching of magnetization with $\vec{I} \parallel \hat{b}$ **(f)** and $\vec{I} \parallel \hat{a}$ **(g)**. The blue, red and white region correspond to unswitched, switched and metastable magnetic states, respectively.

**Fig. 3. Pulse current induced AHE loop shift due to SOT. a,b,** The measured AHE hysteresis loops with $\vec{H} \parallel \hat{z}$ while $500\ \mu s$ current pulses are applied along the *a*-axis **(a)** and *b*-axis **(b)** at each field. **c,d,** The field shift $\mu_0 H_{sh}$ and magnetization reversal field $\mu_0 H_c^{\pm}$as a function of the applied pulse current along the *a*-axis **(c)** and *b*-axis **(d)**. The white, light-grey, and dark-grey regions depict the field-induced, current-assisted, and current-induced magnetization switching regime, respectively.

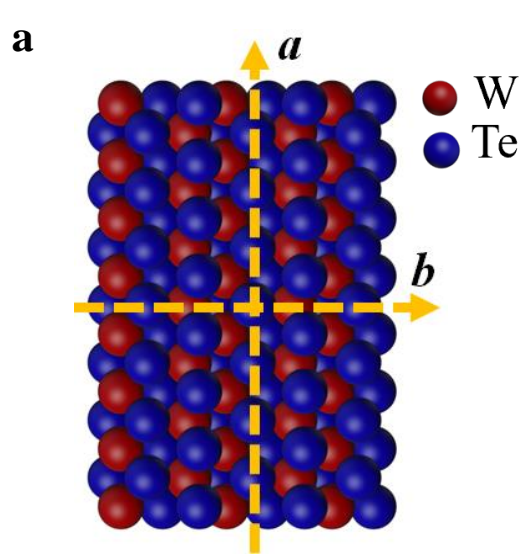
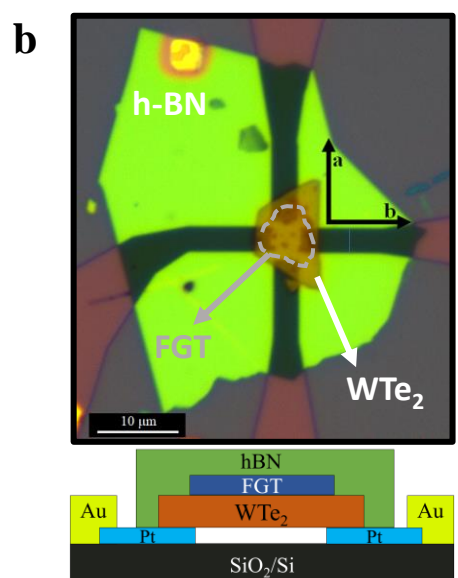
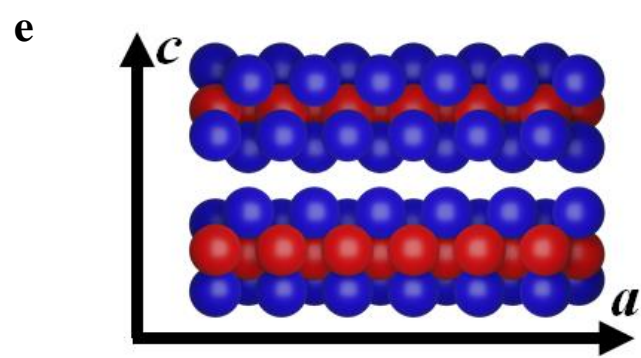
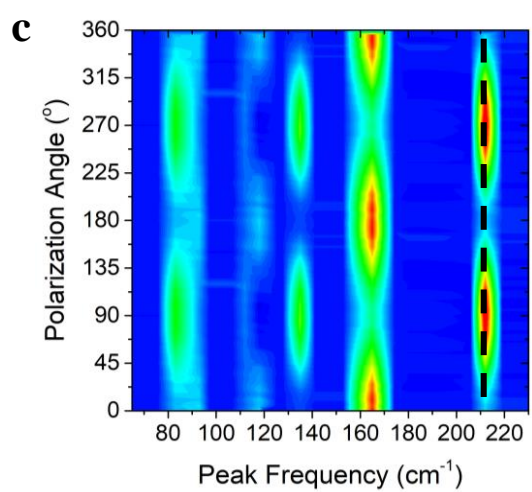
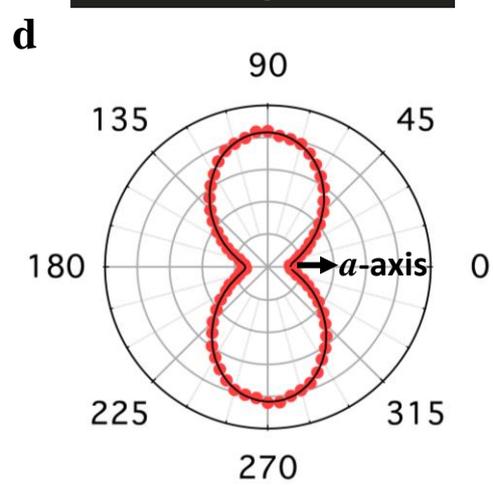
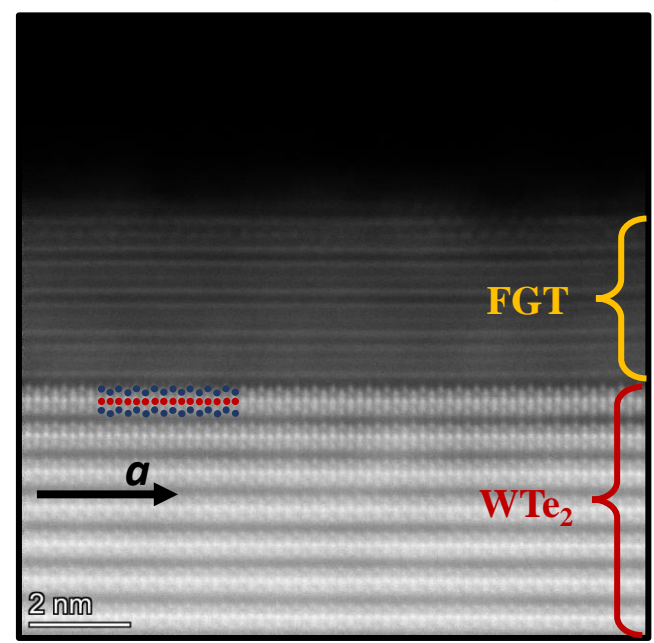

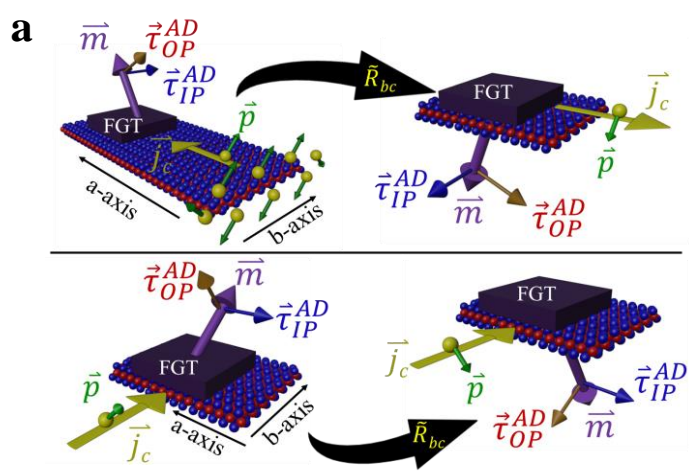
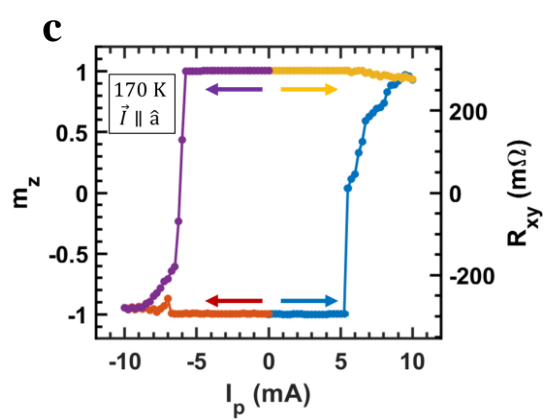
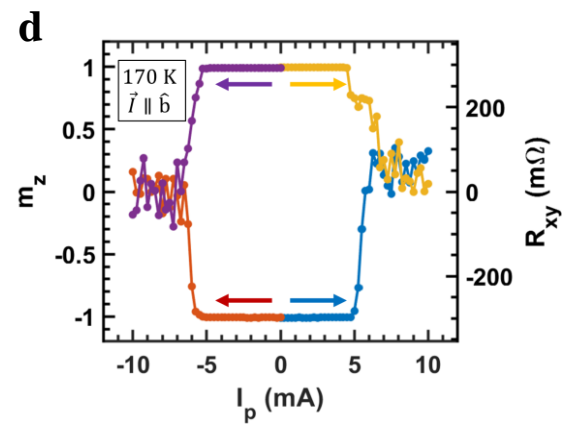
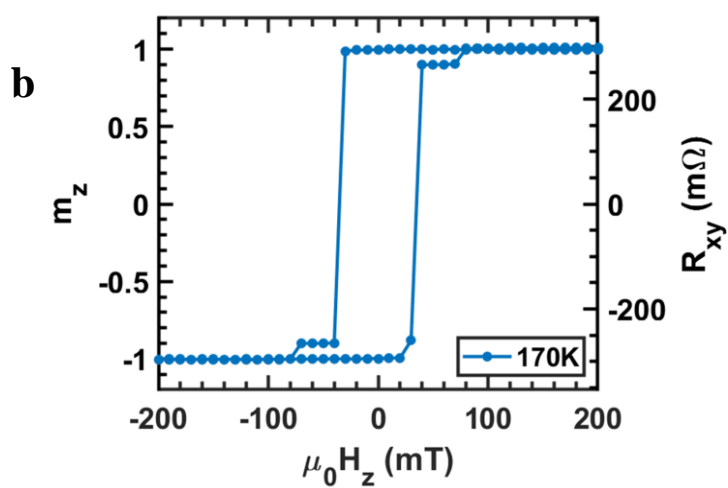
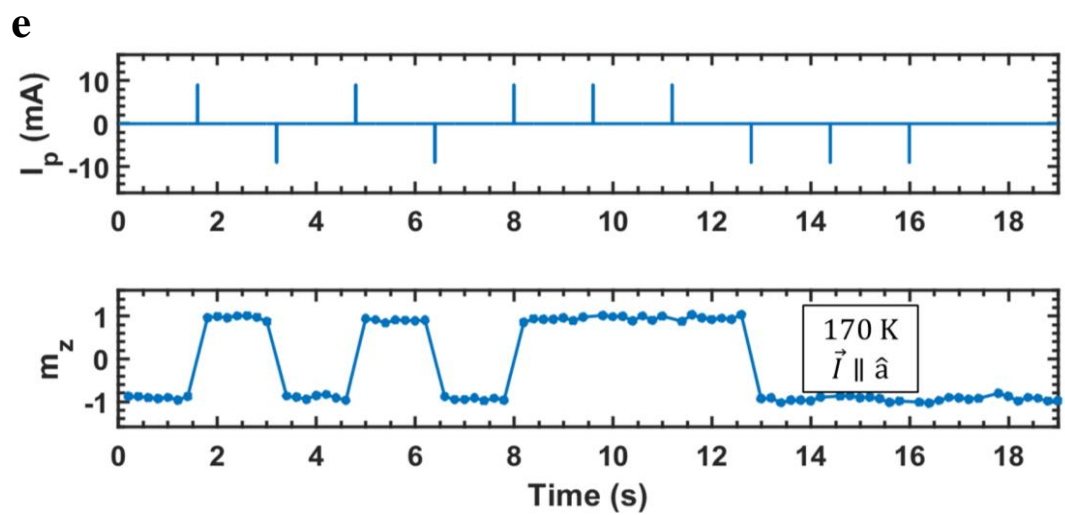
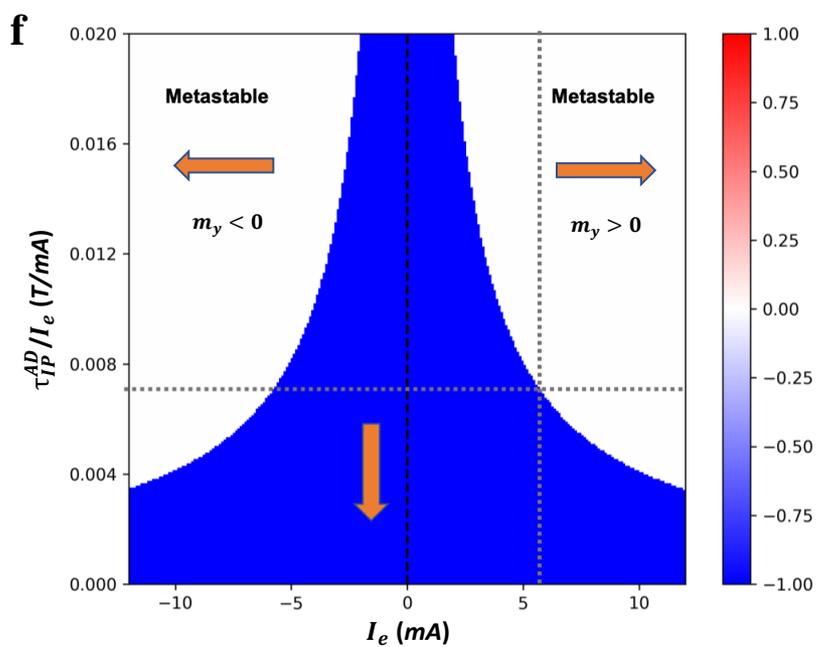
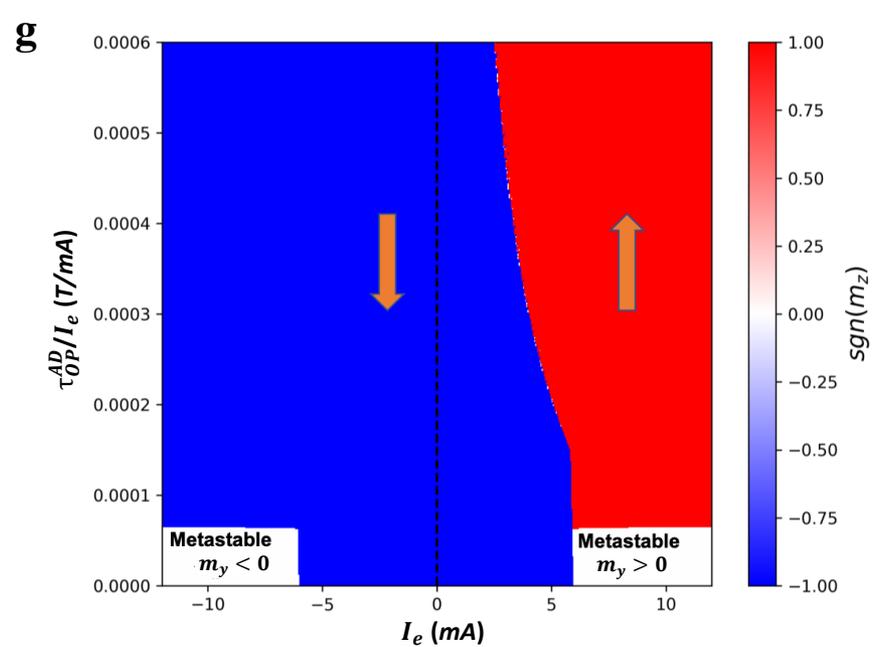

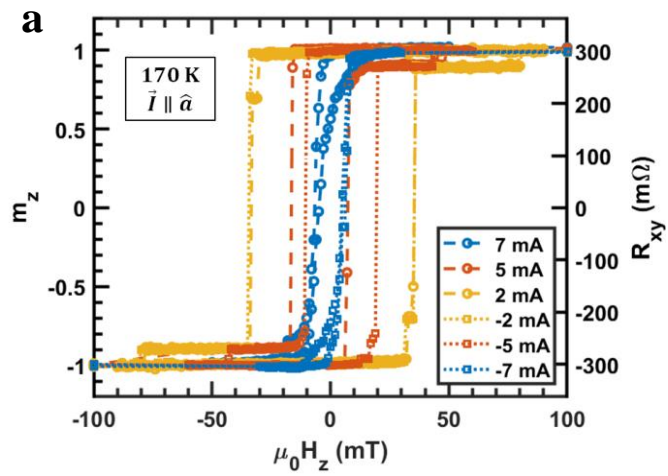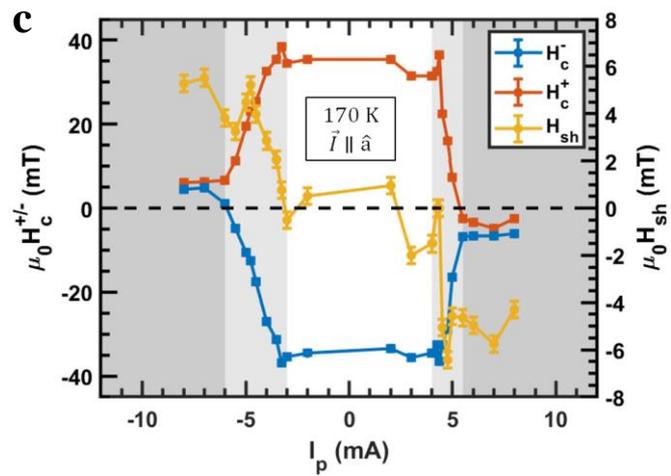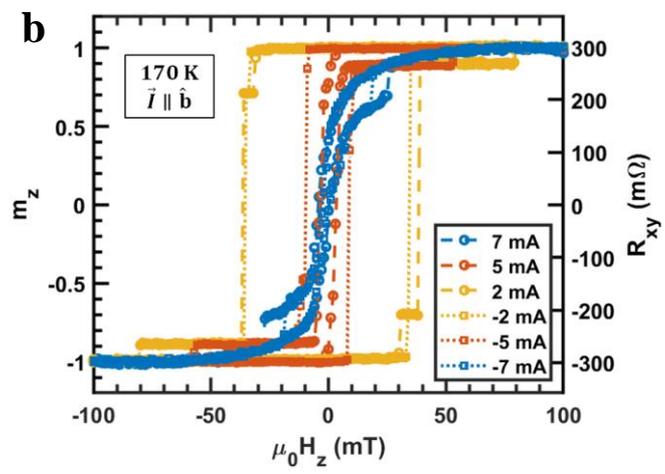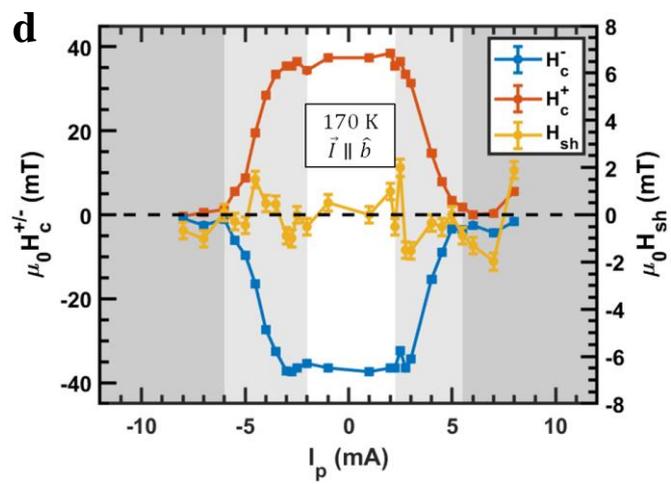

# Contents

**Note 1.  Device Fabrication**

Mechanical exfoliation of WTe$_2$, hBN, and Fe$_{2.78}$GeTe$_2$ (FGT) was performed on separate silicon wafers with 300nm of SiO$_2$ inside of an Ar environment. Flakes were selected through optical investigation through a microscope. WTe$_2$ flakes that have well defined and straight edges were used because the a-axis tends be along them[1]. Prior to heterostructure fabrication, electrodes were defined on a separate Si/SiO$_2$ chip using electron beam lithography and electron beam deposition with a PMMA/MMA bilayer resist. The electrodes were made such that Pt(8 nm) contacted the heterostructure and Cr(5nm)/Au(110nm) contacted the Pt for wire bonding pads. The heterostructure was fabricated using a custom transfer tool in an Ar environment. A transfer slide consisting of a polydimethylsiloxane (PDMS) slab and thin film of polycarbonate (PC) was used for picking up hBN, FGT, and WTe$_2$ in that order and then putting the stack on the Pt electrodes.

**Note 2.   Thickness, Interface, and Crystallographic Orientation of WTe$_2$/FGT Devices**

Three WTe$_2$/FGT van der Waals (vdW) heterostructures were fabricated and named device A, B, and C. All of the results presented in the main paper come from device A. To determine the thickness, crystallographic orientation, and interface quality of the devices, we utilize atomic force microscopy (AFM), polarized Raman spectroscopy, and scanning transmission electron microscopy (STEM).

Device A was made by mechanically stacking exfoliated flakes of WTe$_2$ (25.8 nm)/ FGT (4.1 nm)/ hBN (39.6 nm) on to a pre-written Pt Hall cross, allowing for current to be applied in both the a- and b-axis of WTe$_2$. Device B has a heterostructure of WTe$_2$ (24.4 nm)/ FGT (7.7 nm)/ hBN (30-40 nm), stacked on a pre-written Pt Hall bar, which only allows current to be passed through the b-axis of WTe$_2$. The crystallographic orientation of device A is confirmed by both cross-sectional STEM and polarized Raman, while the axis of orientation of device B is confirmed solely through cross-sectional STEM. Device C consists of a heterostructure of WTe$_2$ (1.44 nm)/ FGT (10.3 nm) resting on a pre-written Pt Hall bar where the relationship between the current channel direction and the crystallographic orientation of WTe$_2$ was not verified. Both device A and B are capped with hBN to avoid oxidation. The results shown in all figures are from device A unless otherwise stated.

For device A, individual flakes and the resulting device can be found in Figure S1a. To characterize the thickness of the components of the device, we utilize AFM in semi-contact mode. The tip scanning direction was set to be near-perpendicular to all steps to acquire the correct thickness. Figure S1b,c, and d show the thickness of Pt, WTe$_2$, and hBN respectively. Arrows in Figure S1a show the location and direction of scanning for each height measured in b-d. The results from AFM are in good agreement with the cross-sectional STEM results (Fig. 1e in the main text).

Polarized Raman spectra were collected in an inVia Renishaw Raman microscope. The device was loaded onto the microscope stage and positioned in such a way that the long edge of the flake (as indicated in Figure S1a) was aligned parallel to the laser polarization ($\theta = 0°$). In this configuration, the incident laser is polarized in the vertical polarization at $\theta = 0°$. The incident laser was rotated from 0 - 360° by 5° increments using a polarization rotator, while an analyzer was set to only allow vertically polarized light to enter the spectrometer. Raman spectra were collected at each polarization with 633 nm excitation, and two accumulations of 10 s. The laser power was set to 1 mW to avoid any damage by heating. Following spectral collection, the (baseline corrected) integrated intensities under each peak were calculated for the contour plot (Fig. 1c in the main text) and polar plots in Figure S2.

For device B, individual flakes can be found in Figure S3a as well as the completed stack on the pre-written electrodes. We utilize cross-sectional STEM to characterize both the crystallographic orientation of $WTe_2$ and the thickness of each component by making the STEM cut along the current channel. As displayed in Figure S3b, the STEM image shows the sample structure: $SiO_2$/Pt/$WTe_2$/FGT/hBN (from bottom to top). Amorphous $WTe_2$ at the Pt/$WTe_2$ interface is due to either oxidation from the suspended region or due to plasma cleaning when preparing the sample for STEM. A zoomed in STEM image at the $WTe_2$/FGT interface (Fig. S3c) shows an atomically sharp interface and that the STEM cut is along the b-axis of $WTe_2$ (and therefore the current direction of device B).

For device C, individual flakes can be found in Figure S4a as well as the completed stack on the pre-written Pt electrodes. Figure S4b and c show the thickness of $WTe_2$ and FGT respectively. The arrows shown in Figure S4a show the location and direction of scanning for each height measured in b and c. We find the thickness of WTe$_2$ to be 1.44 nm, corresponding to bilayer. The thickness of FGT was determined to be 10.3 nm.

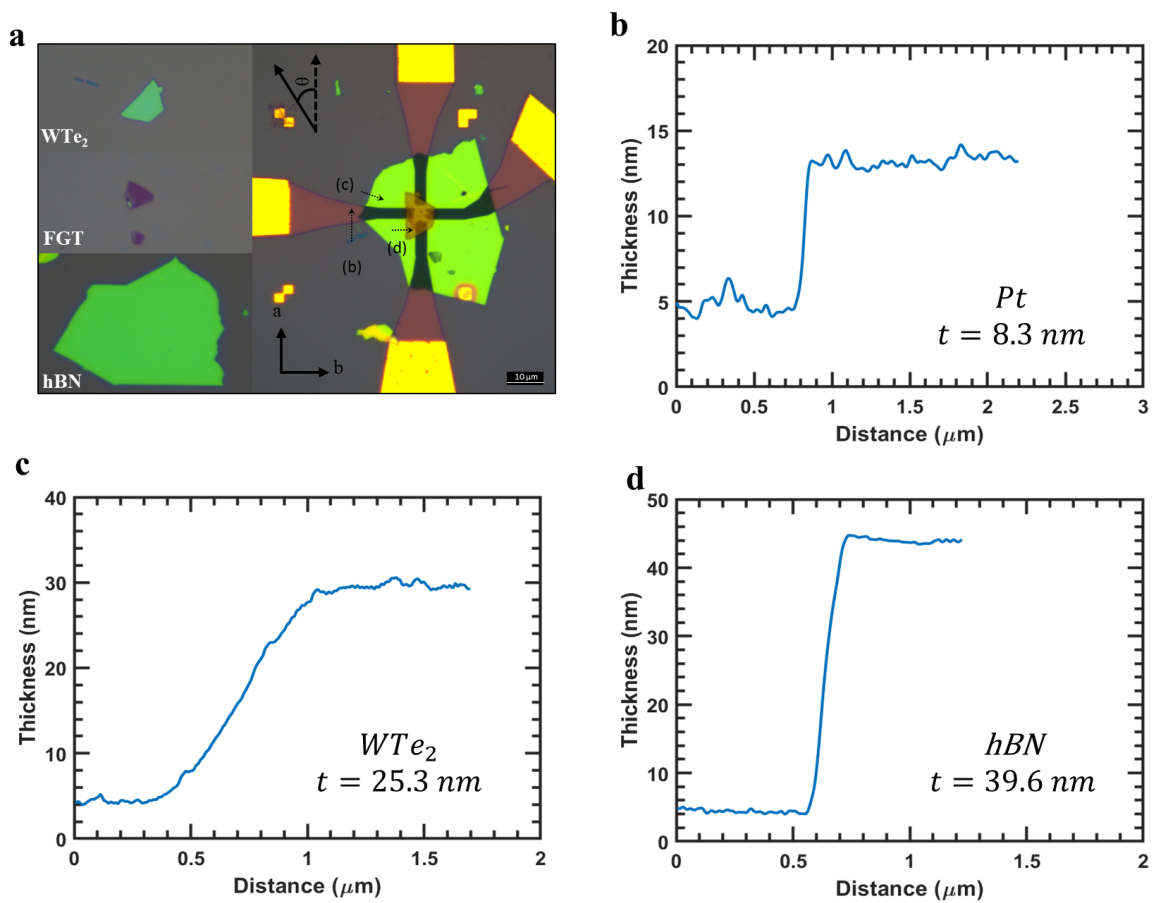

**Figure S1 | The thickness characterization of device A obtained by AFM line scans. a,** the optical images of individual 2D flakes (left) and the complete stack (right), where the dotted arrows show the lines scanned by AFM tip. **b,c,d,** relative thickness as a function of distance obtained by AFM line scan along the dotted arrows, reflecting the thickness of the Pt electrodes (**b**), WTe$_2$ (**c**), and hBN (**d**).

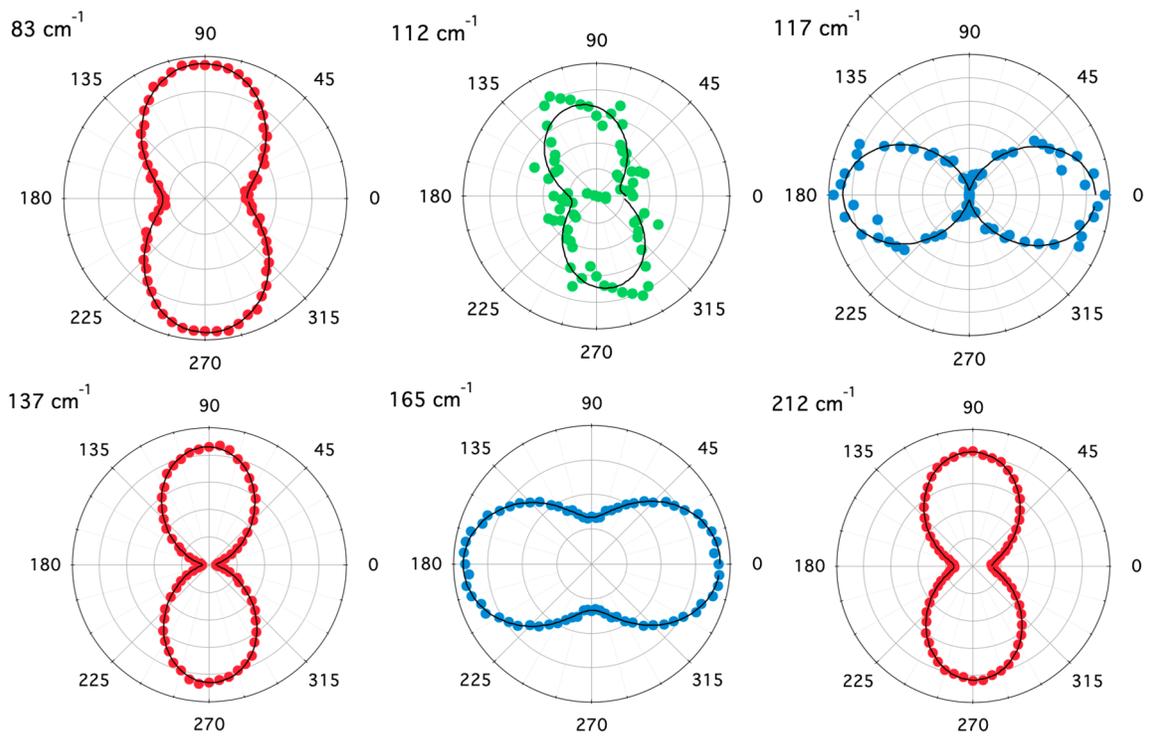

**Figure S2 | Polarized Raman of device A.** Raman data was collected as a function of polarization angle where the $\theta = 0°$ is aligned to the long edge of WTe$_2$ (as shown in figure S1a). Each Raman peak value is displayed on the top left of the corresponding plot. The peaks are individually fitted and the integrated intensity for each peak is plotted for every polarization angle as a polar plot. The data confirms that the a-axis of WTe$_2$ is where $\theta = 0°$.

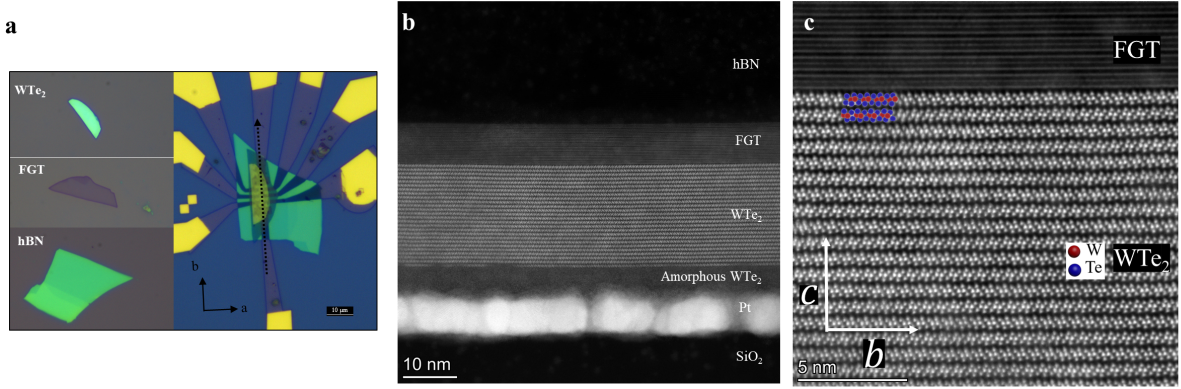

**Figure S3 | The WTe$_2$ crystallographic orientation and thickness characterization of device B. a,** the optical images of individual 2D flakes (left) and the completed stack (right), where the dotted arrow shows the direction of the STEM cut. **b,** The cross-sectional STEM of device B, showing the stack structure and atomically sharp interfaces. **c,** The zoomed in cross-sectional STEM image of the interface of WTe$_2$ and FGT showing that the cut is along the b-axis of WTe$_2$. Our model is over laid to denote the type of atom.

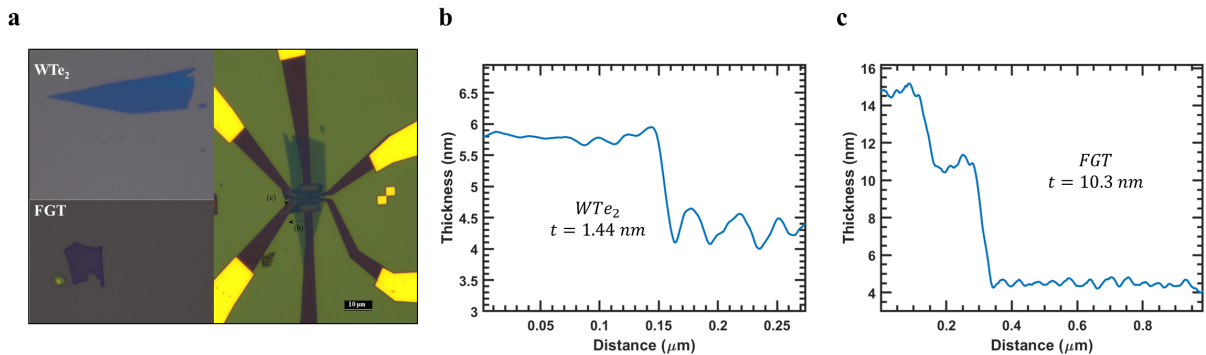

**Figure S4 | The thickness characterization obtained by AFM line scans on device C. a,** Optical images of individual 2D flakes (left) and the complete stack (right), where the dotted arrows show the lines scanned by the AFM tip. **b,c,** relative thickness as a function of distance obtained by AFM line scan along the dotted arrows, reflecting the thickness of WTe$_2$ **(b)** and FGT **(c)**.

**Note 3.   Longitudinal Resistance and Shunt Factor**

To quantify the amount of current shunting through the WTe$_2$, the longitudinal resistance R$_{xx}$ of all devices are measured. First, we define $\hat{x}$ as as the direction of the positive current. Considering the parallel resistor model, the longitudinal resistivity of the device can be written as $\rho_{xx} = \rho_{xx}^{WTe_2}[\frac{t_{WTe_2}}{t} + \frac{\rho_{xx}^{WTe_2}}{\rho_{xx}^{FGT}}\frac{t_{FGT}}{t}]^{-1}$, where $\rho_{xx}^{WTe_2}$ ( $\rho_{xx}^{FGT}$) is the resistivity of WTe$_2$ (FGT), $t_{WTe_2}$ ( $t_{FGT}$) is the thickness of WTe$_2$ (FGT), and $t = t_{WTe_2} + t_{FGT}$. Since WTe$_2$ has a lower resistivity compared to FGT, the total resistivity $\rho_{xx}$ is mostly contributed by WTe$_2$. Especially for device A and B, we expect $\rho_{xx} \approx \rho_{xx}^{WTe_2}$ because $t_{WTe_2} >> t_{FGT}$. Previous reported values of $\rho_{xx}^{FGT}(T)$ have shown rather weak temperature dependence[2–4], while $\rho_{xx}^{WTe_2}(T)$ has shown to have a strong temperature dependence[5].

Since the Hall bar geometry of device B allows for 4-point measurments, we first show the longitudinal resistivity results of device B in Figure S5a. The strong temperature dependence of $\rho_{xx}$ confirms that $\rho_{xx} \approx \rho_{xx}^{WTe_2}$. To estimate $\rho_{xx}^{WTe_2}$, we take the values reported by Z. Fei *et al.*[3] for the resistivity of FGT and use the parallel resistor model to estimate the resistivity of WTe$_2$, We note that our estimated $\rho_{xx}^{WTe_2}$ is very close to the value reported by P. Li *et al.* for WTe$_2$ flakes with similar thickness (35 nm)[5]. We then calculate the shunt factor as a function of temperature, $X(T)$, for both WTe$_2$ and FGT (Fig. S5b), which indicates that most of the current shunts through WTe$_2$ in device B.

As for device A, $R_{xx}$ is measured by the 2-point measurement along the a-axis instead of the 4-point measurement due to the geometry of the Hall cross. The WTe$_2$/FGT heterostructure is assumed as a cuboid with dimensions of 5 $\mu m$ × 3 $\mu m$ × 29.9 $nm$. The measured resistance includes the resistance of the WTe$_2$/FGT heterostructure, Pt electrodes, and contact resistance $R_C$. As shown in Figure S5c, the contribution of the Pt (8 $nm$) electrodes, with total length $\sim$ 39 $\mu m$, is subtracted from the measured total resistance to obtain the resistance of the WTe$_2$/FGT bilayer by considering resistors in series. We use the values reported by M. Alghamdi *et al.* for the resistivity of Pt[2]. The contact resistance is assumed to be independent of temperature and equal to 209 $\Omega$ by matching $\rho_{xx}^{WTe_2}$ between device A and B in the low temperature regime ( T < 50K). The longitudinal resistivity results of device A (Fig. S5d) show minor differences in $\rho_{xx}^{WTe_2}$ with respect to B. Such a difference could be explained by the fact that the current is applied along the a- (b-) axis in device A (B) respectively. We then calculate $X(T)$ for both WTe$_2$ and FGT as a function of temperature (Fig. S5e), which again indicates that most of the current shunts through WTe$_2$. The shunt factor is used to calculate the current density flowing in WTe$_2$ ($J_{WTe_2}$), and hall resistivity $\rho_{xy} = E_y/J_{FGT}$, where $J_{FGT}$ is the current density flowing in FGT.

The longitudinal resistivity results of device C (Fig. S5f) show that WTe$_2$ and FGT almost equally contribute to the total resistivity of the device because $t_{FGT} > t_{WTe_2}$. The estimated $\rho_{xx}^{WTe_2}$ in device C may be lower than the real value, since FGT is less resistive. The Shunt factor of WTe$_2$ and FGT (Fig. S5g) indicates that the current flows almost equally in WTe$_2$ and FGT in device C. Note that the real value of the current flowing in FGT may be larger than the estimated value because of the lower $\rho_{xx}^{FGT}$.

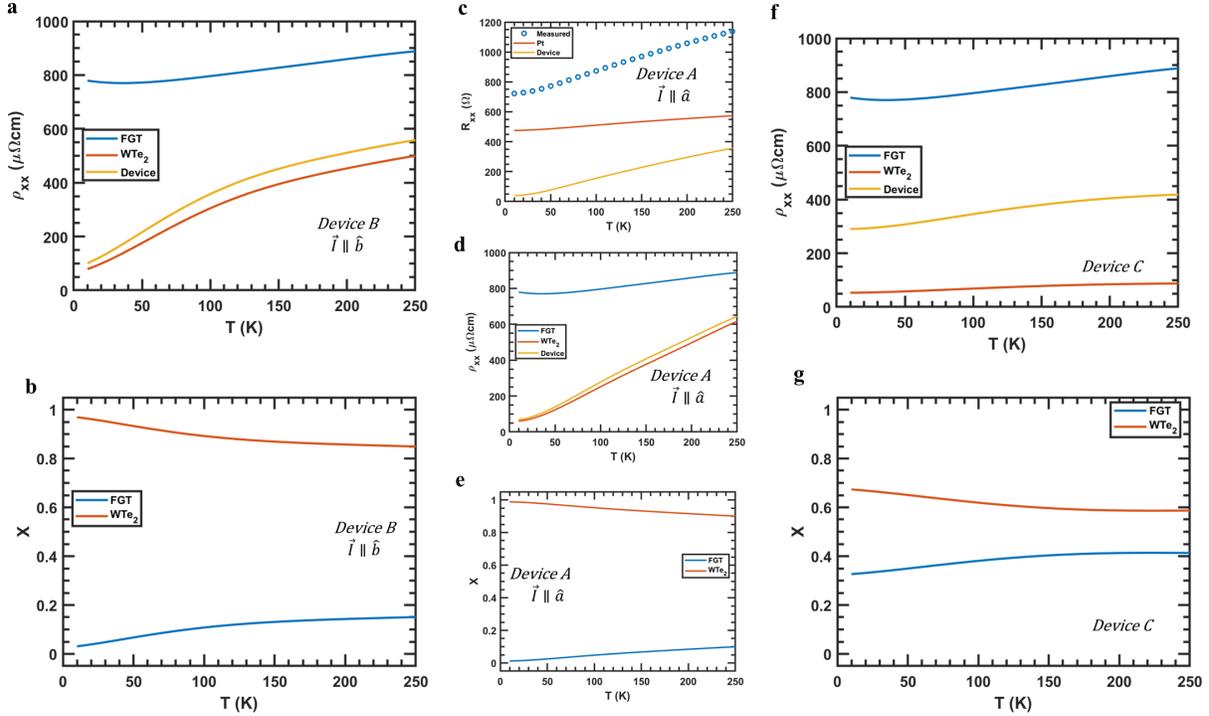

**Figure S5 | The temperature-dependence of longitudinal resistance of WTe$_2$/FGT heterostructures. a,** The estimated longitudinal resistivity $\rho_{xx}$ of the WTe$_2$/FGT bilayer, WTe$_2$ layer, and FGT layer in device B. **b,** The shunt factor as a function of temperature for device B. **c,** the longitudinal resistance $R_{xx}$ of device A as a function of temperature determined by 2-point measurements and the estimated contribution from the 39 $\mu$m-long Pt (8 nm) electrodes. **d,** the estimated longitudinal resistivity $\rho_{xx}$ of the WTe$_2$/FGT bilayer, WTe$_2$ layer, and FGT layer in device A. **e,** The shunt factor as a function of temperature for device A. **f,** the estimated longitudinal resistivity $\rho_{xx}$ of the WTe$_2$/ FGT bilayer, WTe$_2$ layer, and FGT layer in device C. **g,** The shunt factor as a function of temperature for device C.

## Note 4. Temperature Dependence of AHE Hysteresis Loops

For device A, the Hall resistivity, $\rho_{xy}(\mu_0 H_z)$, as a function of the field applied along the easy axis of FGT was measured at various temperatures from 50 K to 220 K (Fig. S6a). These results confirm hard magnet behavior, highly square hysteresis loops, and no indication of the formation of multi-domains. We define $\rho_{AHE}(T)$ as the half-resistivity difference between the two saturated states and plot it as a function of temperature (Fig. S6b). We confirm the existence of a ferromagnetic phase. The antiferromagnetic coupling and a similar peak at $T < T_c$ have been reported in pristine FGT[6,7]. The coercive field $H_c = (H_c^+ - H_c^-)/2$, where $H_c^\pm$ is the positive/negative field for which the magnetization reverses, increases with decreasing temperature (Fig. S6c), and reaches 360 mT at 50 K. We also determined the effective magnetic anisotropic field $\mu_0 H_k \approx 4.12$ T at 170 K by measuring the hard axis AHE loop (Supplementary Note 6), which indicates strong perpendicular magnetic anisotropy. The coercive field $H_c$ is much smaller than $H_k$, indicating that the switching is through incoherent magnetization reversal.

We also perform field-dependent Hall resistivity measurements at various temperatures on device B and C and obtain very similar results (Fig. S7,8). The trend of the measured $\rho_{AHE}$ for device C is close to the trend reported in Pt/FGT systems[2]. The Curie temperature of device B and C are higher than device A due to the increased FGT thickness, which agrees with the previously reported thickness dependent study of the magnetic property of FGT flakes[3].

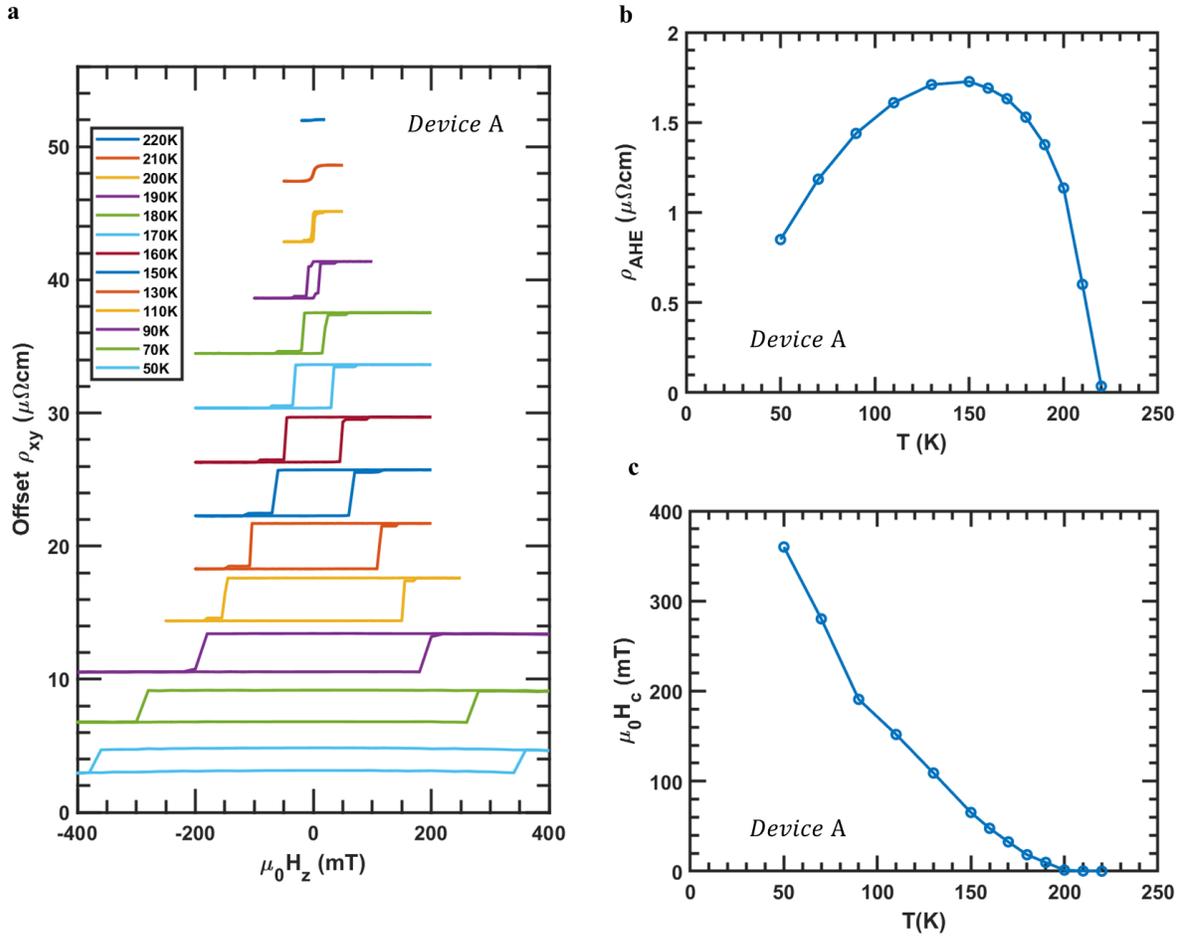

**Figure S6 | The temperature-dependent magnetic property of device A. a,** The transverse resistivity $\rho_{xy}$ as a function of applied field from 50 K to 220 K. **b,** The Anomalous Hall resistivity $\rho_{AHE}$ as a function of temperature. **c,** The coercive field $\mu_0 H_c$ as a function of temperature.

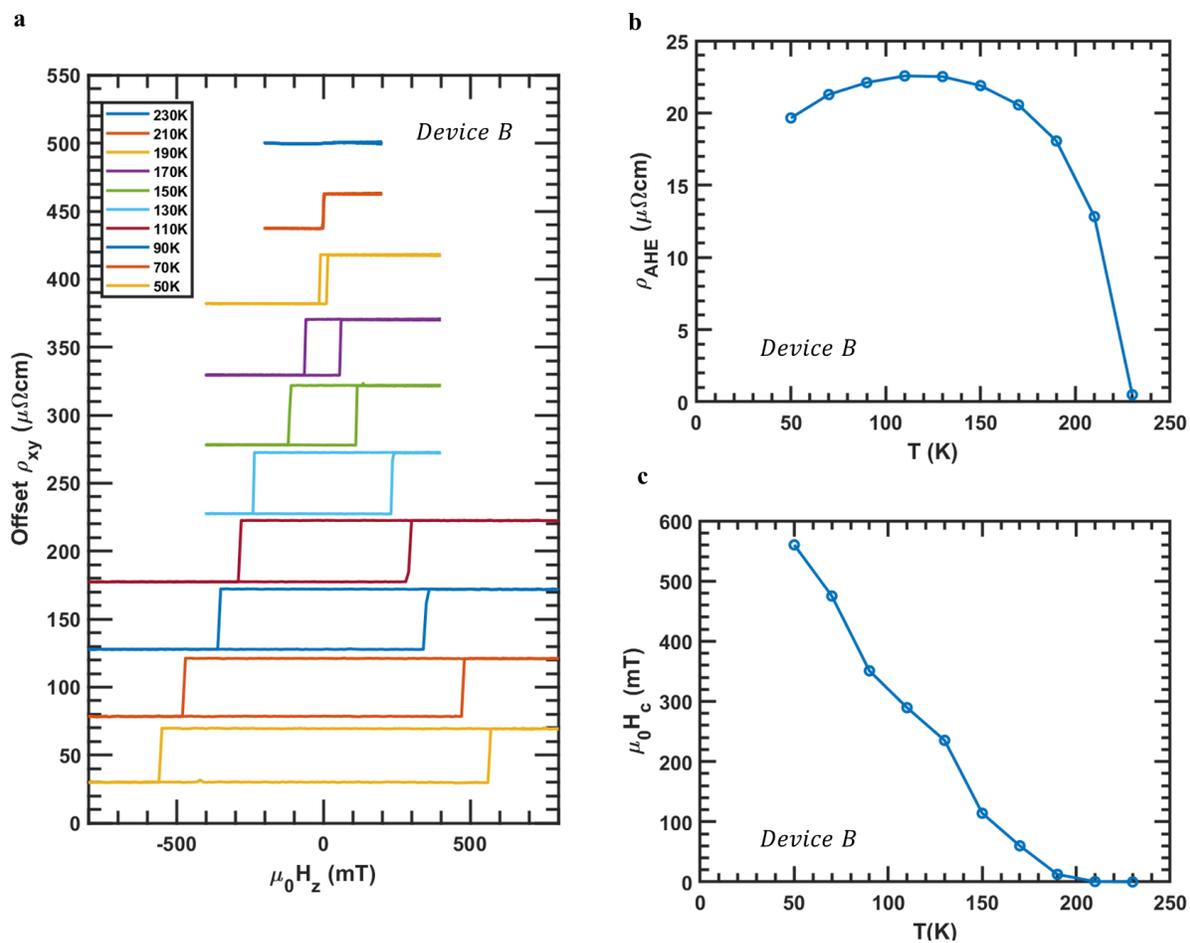

**Figure S7 | The temperature-dependent magnetic property of device B. a,** The transverse resistivity $\rho_{xy}$ as a function of applied field from 50 K to 230 K. **b,** The Anomalous Hall resistivity $\rho_{AHE}$ as a function of temperature. **c,** The coercive field $\mu_0 H_c$ as a function of temperature.

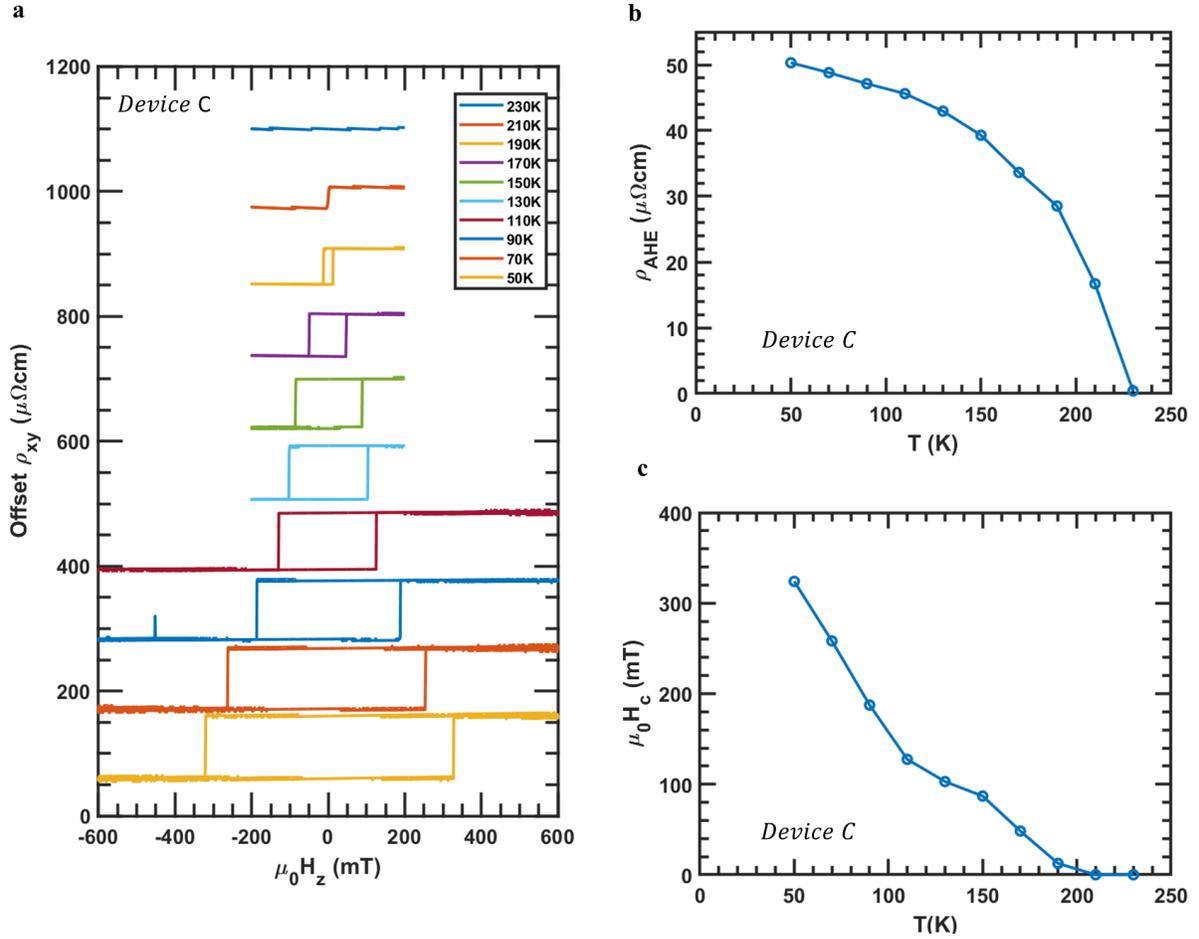

**Figure S8 | The temperature-dependent magnetic property of device C. a,** The transverse resistivity $\rho_{xy}$ as a function of applied field from 50 K to 230 K. **b,** The Anomalous Hall resistivity $\rho_{AHE}$ as a function of temperature. **c,** The coercive field $\mu_0 H_c$ as a function of temperature.

**Note 5. The Pulse Current-Induced Joule Heating and Reduction of Coercive Field**

Current-induced Joule heating has been reported in WTe$_2$/FGT vdW heterostructures[8] and Pt/FGT[4] in similar measurements, which could potentially assist the SOT-induced switching. To examine the impact of Joule heating in device A, due to the limitation of having a hall cross geometry, we utilize the longitudinal component measured in transverse resistance. The measured raw transverse resistance, $R_{xy,raw}(H_z) = R_{OHE}H_z + R_{AHE}m_z(H_z) + R_L$, is a function of the perpendicular field $H_z$ (Fig. S9a), where $R_{OHE}$ is the ordinary Hall resistance, $R_{AHE}$ is the anomalous Hall resistance, $R_L$ is the longitudinal resistance, and $m_z$ is the normalized perpendicular magnetization. By fitting the positive (negative) saturated branch with a linear function, we determine $R_{OHE}$ and define the positive (negative) resistance $R_{pos}(R_{neg})$ as the raw transverse resistance of positive (negative) saturated states at zero field. We can then calculate $R_{AHE} = (R_{pos} - R_{neg})/2$ and $R_L = (R_{pos} + R_{neg})/2$ as a function of temperature by measuring the AHE loop at various temperatures (Fig. S9b). The separation between $R_{pos}$ and $R_{neg}$ is clearly due to AHE and $R_L$ comes from the longitudinal resistance which increases mostly-linearly with increasing temperature. The change of resistance per Kelvin is $\alpha = 2.6 \pm 0.2 \, m\Omega/K$, determined by a linear fit to $R_L(T)$. During the SOT switching measurements, $R_{xy}$ is measured a few seconds after the pulse current is injected. For all the deterministic switching measurements conducted, we observe that $R_{xy}$ always returns to the same point with deviations much smaller than 2.6 $m\Omega$ (a temperature difference of 1 K) after completing a full loop. This shows that there is no substantial heat accumulation in our device.

From the results of the AHE loop shift measurements, we can obtain $R_L(I_p)$ and $R_{AHE}(I_p)$ as a function of pulse current, $I_p$, in the same way. To translate $R_L$ to the actual temperature $T_{JH}$ elevated by Joule heating, we consider $T_{JH}(I_p) = \Delta T(I_p) + T_{sp}$, where $\Delta$T is the temperature difference induced by the applied current pulse and $T_{sp}$ is the set point temperature of the controller. We first use $T(R_L)$ obtained in Fig. S9b to convert $R_L(I_p)$ to $T_{JH}(I_p)$. We find that $T_{JH}(I_p)$ has a clear $I_p^2$ dependence, which indicates Joule heating. Compared to the resistance obtained by delta mode, the resistance obtained by pulse delta mode has a resistance offset, so we use the quadratic fit result of $T_{JH}(I_p)$ to enforce $T_{JH}(I_p = 0) = T_{sp}$. As shown in Figure S9c and e, the $I_p$ dependence of $T_{JH}$ when $\vec{I} \parallel \hat{a}$ and $\vec{I} \parallel \hat{b}$ both show that $T_{JH}$ could surpass 240 K at $I_p = 8mA$, which corresponds to $J_p \approx 9.44 \times 10^{10} A/m^2$.

The fact that $T_{JH} > T_c$ may be surprising, however, it should be noted that the temperature of FGT may not be the same as the temperature of WTe$_2$. Since most of the current is shunting through WTe$_2$, $R_L$ may not be a good indicator for the elevated temperature of FGT ($T_{JH}^{FGT}$). To verify this, we use $T(R_{AHE})$ (obtained in Fig. S9b) to convert $R_{AHE}(I_p)$ to $T_{JH}^{FGT}(I_p)$ and find that $T_{JH}^{FGT}(I_p)$ has a similar trend but is roughly 30 K below $T_{JH}(I_p)$. Points in the low current regime for the linear interpolation of $T_{JH}^{FGT}(I_p)$ are omitted because $R_{AHE}$ has a weak temperature dependence near 170K as shown in Figure S9g. We also note that the $T_{JH}$ is higher at the same $I_p$ when $\vec{I} \parallel \hat{b}$, which may be due to slight differences in resistance when current is applied along different axes.

To link the Joule heating to the observed reduction of $H_c$ induced by current injection, we plot $H_c(I_p)$ when $\vec{I} \parallel \hat{a}$ and $\vec{I} \parallel \hat{b}$ in Figure S9d and f. The onset of the reduction in $H_c$ starts close to where $T_{JH}^{FGT}$ starts to rise, indicating that the reduction of $H_c$ is related to Joule heating. Because the spin torques generated by the current may also reduce $H_c$, the reduction of $H_c$ may be the consequence of Joule heating and current-induced spin torque. Note that the peak in $H_c$ cannot be explained by Joule heating, which is the signature of magnetic excitations due to spin current induced steady processional states. These results indicate that the current-induced SOT switching in the WTe$_2$/FGT heterostructure is similar to the thermal-assisted switching, where the current raises the temperature of the ferromagnetic layer in order to lower the energy barrier between two stable states.

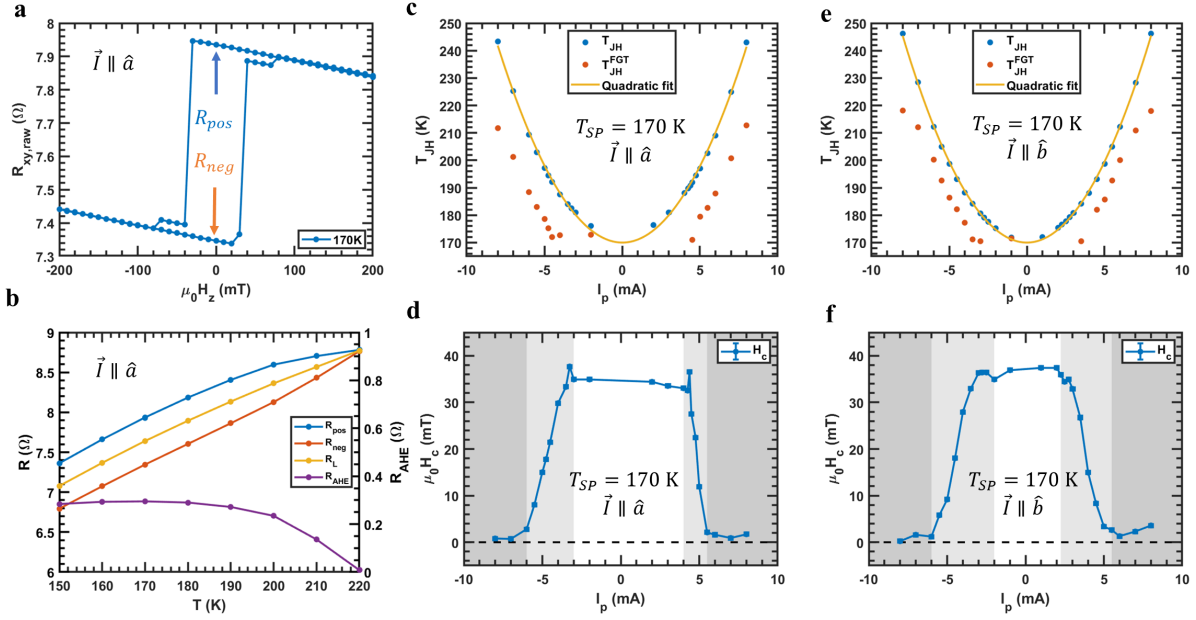

**Figure S9 | The current-induced coercive field reduction and joule heating. a,** The AHE hysteresis loop as a function of the perpendicular magnetic field $H_z$ at 170 K without the background removed, where $R_{pos}$ ($R_{neg}$) is defined as the resistance measured at the positive (negative) state at zero field. **b,** The longitudinal component $R_L$ and $R_{AHE}$ extracted from the measured transverse resistance $R_{xy}$ as a function of temperature plotted with $R_{pos}$ and $R_{neg}$. $R_{AHE}$ is plotted on the right axis. **c,e,** The estimated temperature raised by joule heating, $T_{JH}$, and $T_{JH}^{FGT}$ as a function of $I_p$ applied along the a-axis **(c)** and b-axis **(e)**. **d,f,** The $I_p$ dependence of $H_c$ measured with $\vec{H}\|\hat{z}$ while the current pulse is applied along the a-axis **(d)** and b-axis **(f)**.

## Note 6. Effective Magnetic Anisotropy H$_k$ and in-Plane SOT in WTe$_2$/ FGT Heterostructure

The in-plane current in WTe$_2$ is expected to induce a strong in-plane spin polarization, which could exert in-plane field-like and antidamping torques to the magnetic moment in FGT. To examine the strength of current-induced in-plane torque, we apply an external field $H$ tilted away from the x- and y-axis towards the z-axis by a small angle $\beta = 5°$ to measure the effective field generated by the current. This measurement is similar to the one performed by Liu *et al.* in a conventional SOT system[9]. We assume coherent magnetization rotation and measure the normalized perpendicular magnetization $m_z = Cos\theta = R_{xy}/R_{AHE}$ at each field, where $\theta$ is the spherical polar angle of the magnetization and $\hat{z}$ is the out-of-plane direction. The resistance is measured by averaging 10 consecutive 500 $\mu s$ current pulses in pulse delta mode to avoid heat accumulation. Note that the effect of OHE and $R_{AHE}$ is determined independently by measuring the AHE loop with $\vec{H} \parallel \hat{z}$.

When current is applied along the x-axis, the in-plane spin polarization is in the y-direction, which could potentially generate effective antidamping and field-like fields denoted as $H_{IP}^{AD}$ and $H_{IP}^{FL}$ respectively. Since $H_{IP}^{AD}$ is collinear to $\hat{m} \times \hat{y}$ and $H_{IP}^{FL}$ is collinear to $\hat{y}$, the AHE loop will be modulated by $H_{IP}^{AD}$ ($H_{IP}^{FL}$) when $\vec{H} \parallel \hat{x}$ ($\vec{H} \parallel \hat{y}$). The external field needed to tilt $\hat{m}$ away from the z-axis by $\theta$, when $\pm I_p$ is applied, can be written as $H_{\pm}(\theta) = [H_k Sin\theta Cos\theta \mp H_{IP}^{AD(FL)}]/Cos(\theta + \beta)$, derived by Liu *et al.*[9]. We define the even part $H_{even}(\theta) = [H_+(\theta) + H_-(\theta)]/2 = H_k Sin\theta Cos\theta/Cos(\theta + \beta)$, and the odd part $H_{odd}(\theta) = [H_-(\theta) - H_+(\theta)]/2 = H_{IP}^{AD(FL)}/Cos(\theta + \beta)$. Therefore, by calculating $H_{even}(\theta)$ and $H_{odd}(\theta)$ from the measured AHE loops, $H_k$ and $H_{IP}^{AD(FL)}$ can be obtained by a one parameter fit. We first measure $H_k$ at 170 K in delta mode with a small current of 50 $\mu$A, at which $H_{IP}^{AD}$, $H_{IP}^{FL}$, and Joule heating are negligible. As shown in Figure S10, we determine $H_k \approx 4.14$ T at 170 K, which is close to the reported value by Alghamdi *et al.* at 180 K[2].

To quantify $H_{IP}^{AD}$, we perform AHE loop measurements by applying an external field $\vec{H} \parallel \hat{y}$ for both $\vec{I} \parallel \hat{a}$ and $\vec{I} \parallel \hat{b}$. As shown in Figure S11a and b, the clear difference in $m_z(H)$ between $I_p = +7\,\text{mA}$ and $I_p = -7\,\text{mA}$ indicates strong $H_{IP}^{AD}$. The points at lower fields are excluded because the magnetization reversal occurs in the low field regime. As shown in Figure S11c and d by fitting $H_{even}(\theta)$ and $H_{odd}(\theta)$, we determine $H_{IP}^{AD} \approx 28\,\text{mT}$ and $H_k \approx 2.03\,\text{T}$ when $\vec{I} \parallel \hat{a}$, and $H_{IP}^{AD} \approx 32\,\text{mT}$ and $H_k \approx 1.4\,\text{T}$ when $\vec{I} \parallel \hat{b}$. Strikingly, $H_{IP}^{AD}/J_p \approx 36\,\frac{mT}{10^{11}A/m^2}$ in our device is about 20 times larger than $H_{IP}^{AD}/J_p \approx 1.7\,\frac{mT}{10^{11}A/m^2}$, reported in the conventional SOT system of Pt/Co/AlO$_x$[9]. The observation of strong in-plane antidamping torques is in good agreement with the theoretically predicted high spin-hall angle $\theta_{SHE} = -0.54$ in WTe$_2$[10], and the experimental results from spin-torque ferromagnetic resonance[11,12]. The significant decrease of $H_k$ is due to Joule heating and the difference in $H_k$ between $\vec{I} \parallel \hat{a}$ and $\vec{I} \parallel \hat{b}$ is due to the difference in $T_{JH}$ as mentioned in Note 4.

To quantify the field-like case, $H_{IP}^{FL}$, we measure AHE loops by applying an external field $\vec{H} \parallel \hat{y}$ for both $\vec{I} \parallel \hat{a}$ and $\vec{I} \parallel \hat{b}$. We observe no substantial difference in $m_z(H)$ between $I_p = +7$ mA and $I_p = -7$mA (Fig. S12a and b), showing weak $H_{IP}^{FL}$. As shown in Figure S12c and d, we determine $H_{IP}^{FL} \approx -2.2\,\text{mT}$ and $H_k \approx 1.81\,\text{T}$ when $\vec{I} \parallel \hat{a}$ and $H_{IP}^{FL} \approx -3.7\,\text{mT}$ and $H_k \approx 1.29\,\text{T}$ when $\vec{I} \parallel \hat{b}$. However, we should also consider the Oersted effect, which could generate an Oersted field $\mu_0 H_{Oe} \approx \frac{\mu_0 t_{WTe_2}}{2} J \approx 1.3\,\text{mT}$, pointing towards $\mp \hat{y}$ at $I_p = \pm 7\,mA$. Therefore, we conclude that the observed in-plane field-like field is likely due to the Oersted effect. Furthermore, no significant $H_{IP}^{FL}$ is observed, which is again in good agreement with the experimental results from spin-torque ferromagnetic resonance measurements[11,12].

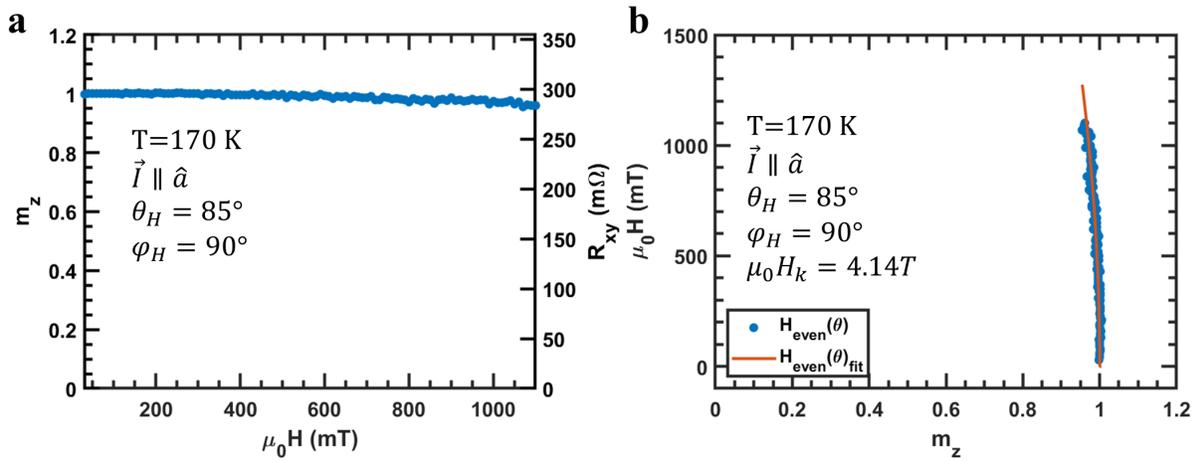

**Figure S10 | The determination of effective magnetic anisotropy. a,** The AHE hysteresis loop as a function of field applied in yz plane but tilted 5° away from the y-axis. **b,** Points: the measured value of H($m_z$) obtained from **(a)**. Line: The fit curve obtained by one parameter fit, which yields $\mu_0 H_k = 4.14$ T.

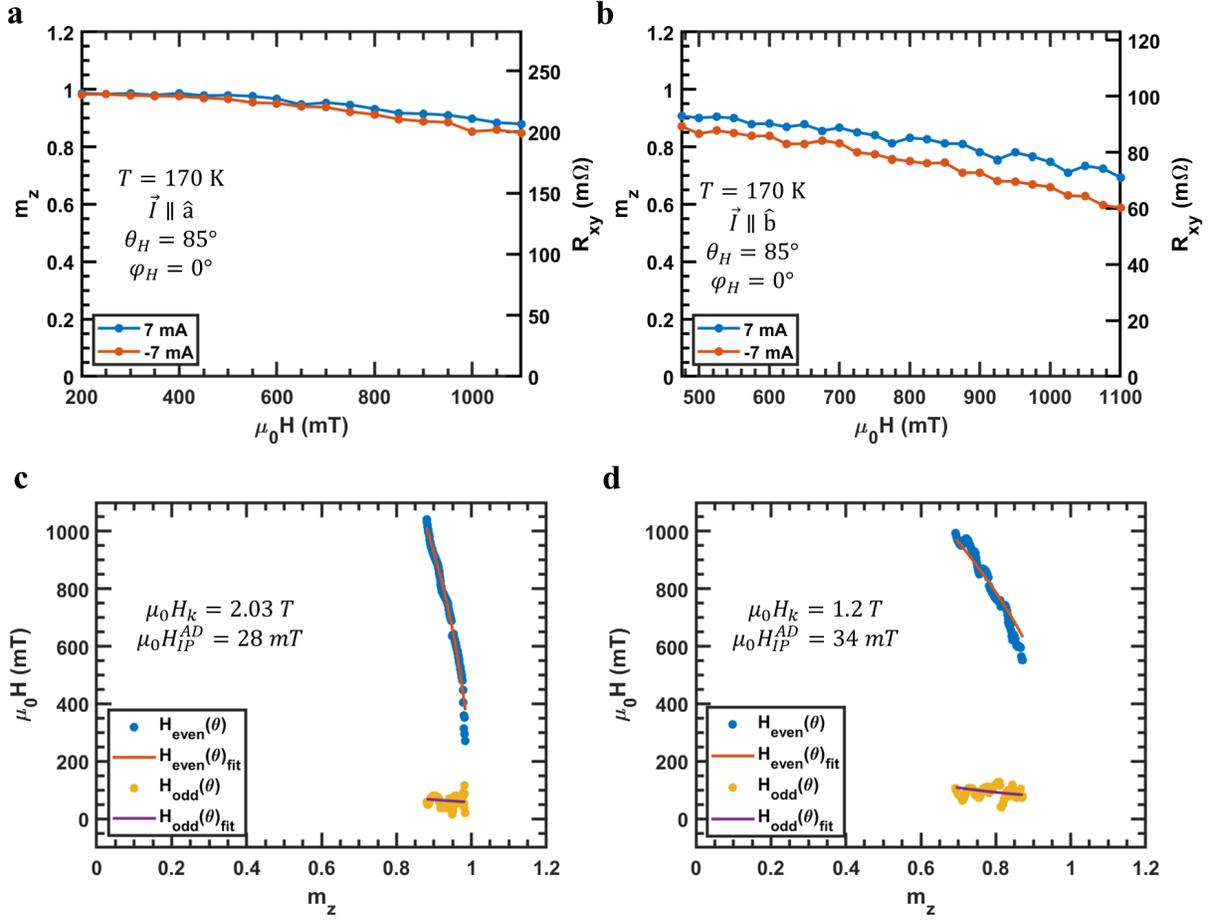

**Figure S11 | The in-plane antidamping effective field in WTe$_2$/FGT. a,b,** The perpendicular magnetization is measured when the field is applied in xz plane but tilted +5° away from x-axis while current applied along the a-axis **(a)** and b-axis **(b)**. **c,d** Points: The measured value of $H_{even}(m_z)$ and $H_{odd}(m_z)$ obtained from **(a)** and **(b)**, respectively. Lines: The fit curve obtained by one parameter fit to determine $H_k$ and $H_{IP}^{AD}$.

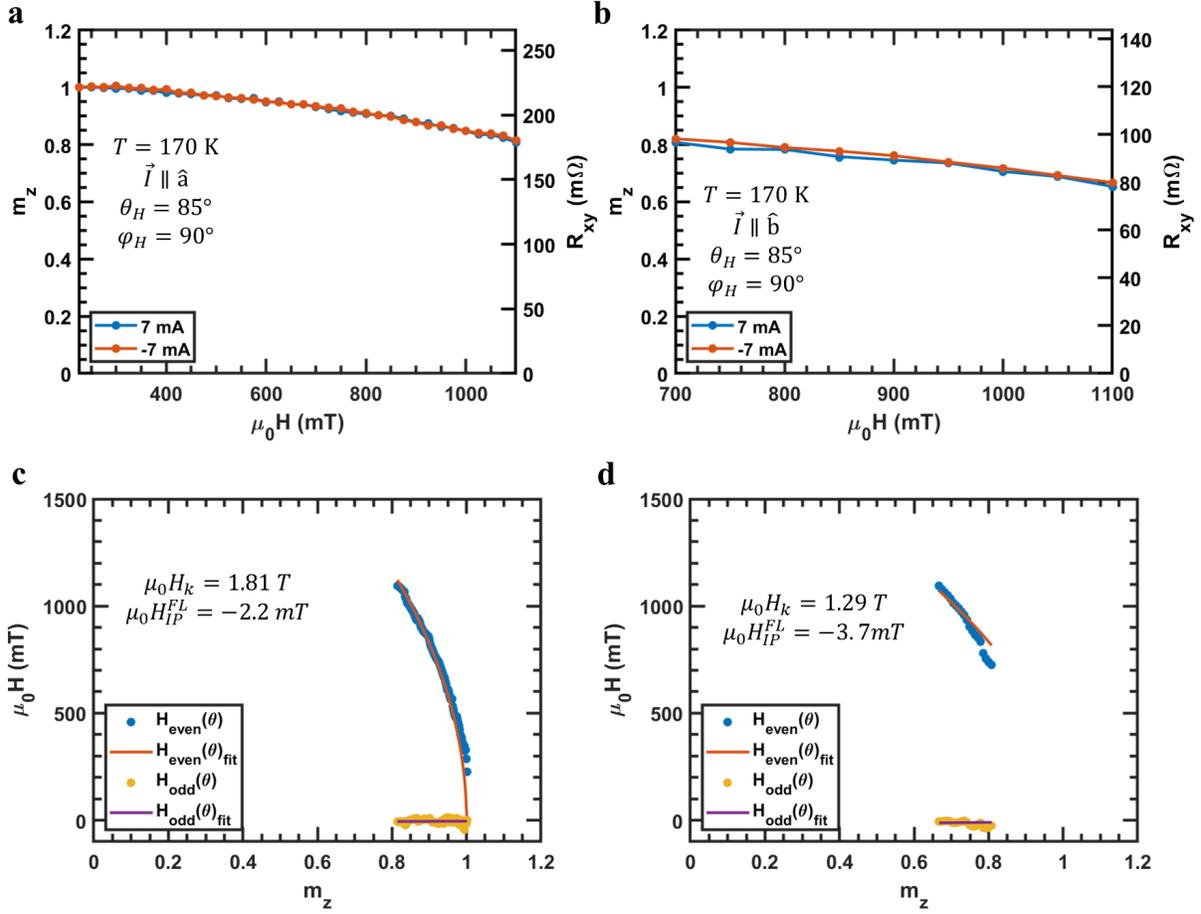

**Figure S12 | The in-plane field-like effective field in WTe$_2$/FGT. a,b,** the perpendicular magnetization is measured when the field is applied in yz plane but tilted +5° away from y-axis while current applied along the a-axis **(a)** and b-axis **(b)**. **c,d,** Points: The measured value of $H_{even}(m_z)$ and $H_{odd}(m_z)$ obtained from **(a)** and **(b)**, respectively. Lines: The fit curve obtained by one parameter fit to determine $H_k$ and $H_{IP}^{FL}$.

**Note 7. Current-Induced SOT Switching in the Presence of an in-Plane Field**

In conventional heavy metal systems, an in-plane field $H_x$ applied collinear to the current is required to achieve deterministic switching due to the absence of $\mathcal{T}_{OP}^{AD}$, which should be a similar scenario when $\vec{I} \parallel \hat{b}$. Therefore, we perform current-induced SOT switching in the presence of $H_x$ to test both the robustness of switching when $\vec{I} \parallel \hat{a}$ and the response of $H_x$ when $\vec{I} \parallel \hat{b}$. The results when $H_x = 0$ mT (Fig. S13a and d) show clear deterministic switching when $\vec{I} \parallel \hat{a}$, while demagnetized states occur at high current values when $\vec{I} \parallel \hat{b}$ due to Joule heating and in-plane torques that tend to form multi-domains. As shown in Figure S13a-c, the applied field of $\pm 100$ mT has little effect on the SOT switching when $\vec{I} \parallel \hat{a}$. This proves that the observed deterministic switching is not the same as SOT switching induced by in-plane antidamping torques which could be assisted by a stray field.

When $\vec{I} \parallel \hat{b}$ with $H_x = \pm 100$ mT, we find that the negative (positive) terminal state at high positive currents moves down (up) (Fig. S13d-f). However, it is difficult to distinguish if the difference is contributed by the change of response to antidamping torques that depend on the sign of $H_x$ or by the $H_z$ component from the small magnetic field misalignment. The formation of multi-domains due to the increase of temperature at high current values also complicates the analysis.

The other signature of switching induced by in-plane antidamping torques is the chirality reversal in the switching hysteresis loop when $H_x$ reverses. However, we do not see clear chirality reversal in device A when $\vec{I} \parallel \hat{b}$, as shown in Fig. S13a and b. The loops are not centered due to the slight misalignment of the magnetic field and the unsaturated magnetization is due to the formation of multi-domains, as observed in Pt/FGT systems[2,4]. It is interesting that the formation of such domains has little influence on the terminal states when $\vec{I} \parallel \hat{a}$. In order to ex-tract details of how the terminal states are achieved by in-plane antidamping torques at $T \approx T_c$ time-resolved domain imaging techniques are needed.

We also perform SOT switching in device B at various fields, including switching at zero field. As shown in Figure S14a, the switching chirality is counter-clockwise when $\mu_0 H_x = +100$ mT. While when $\mu_0 H_x = -100$ mT (Fig. S14b), the chirality reverses to clockwise. The chirality reversal in device B is clear and consistent with switching induced by in-plane antidamping torque. We plot SOT switching results at various fields in Figure S14c and d. For sufficient $|\mu_0 H_x| \geq 20$ mT, the chirality of the loop is determined entirely by the sign of $H_x$, as expected for field-assisted SOT switching. We found that the switching is inefficient at zero field and the chirality is the same as when $H_x < 0$ and the switching is optimized at $|\mu_0 H_x| = 50 - 100$ mT. The switching at zero field may be explained by stray fields, a small $\tau_{OP}^{AD}$ induced by a slight misalignment of the current channel, and (or) current spreading. At higher fields, the loop is narrowed with increasing $|H_x|$, indicating the formation of multi-domains.

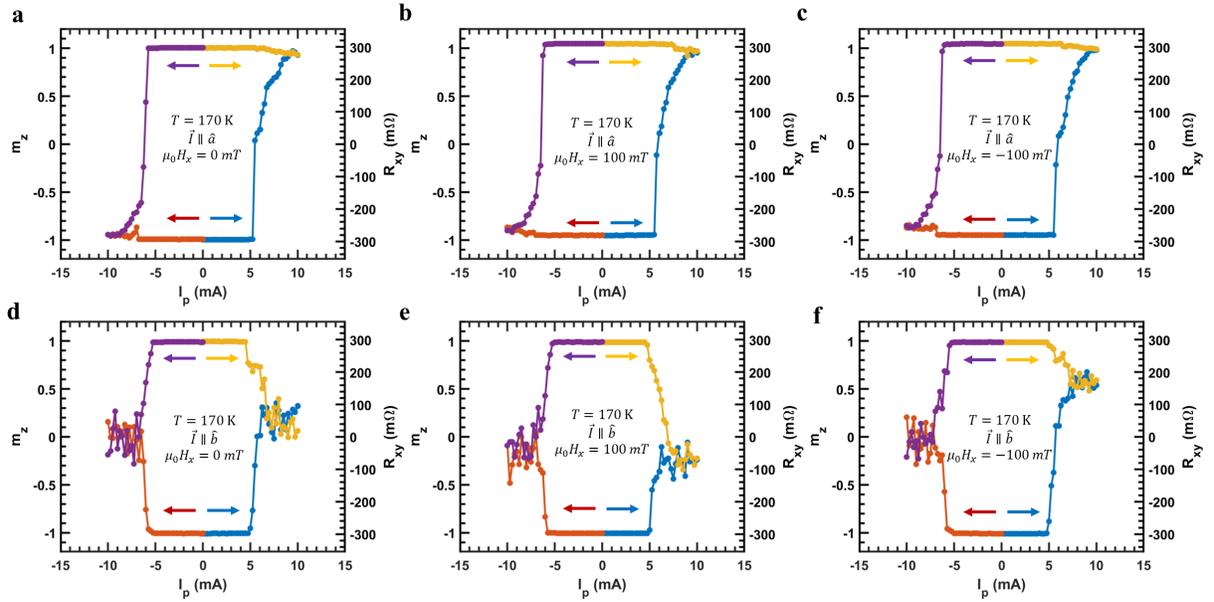

**Figure S13 | The pulse-induced SOT switching of device A with and without field $H_x$ applied along the current.** For all measurements, the magnetization is initialized to $m_z = \pm 1$ prior to the application of the current. Starting from zero, current is gradually pulsed towards the negative or positive direction indicated by arrows. **a-f,** The pulse-induced SOT switching at 190 K while $\vec{I} \parallel \hat{a}$ **(a-c)** and $\vec{I} \parallel \hat{b}$ **(d-f)** with $\mu_0 H_x$= 0 mT **(a,d)**, 100 mT **(b,e)**, and -100 mT **(c,f)** applied collinear to the current direction.

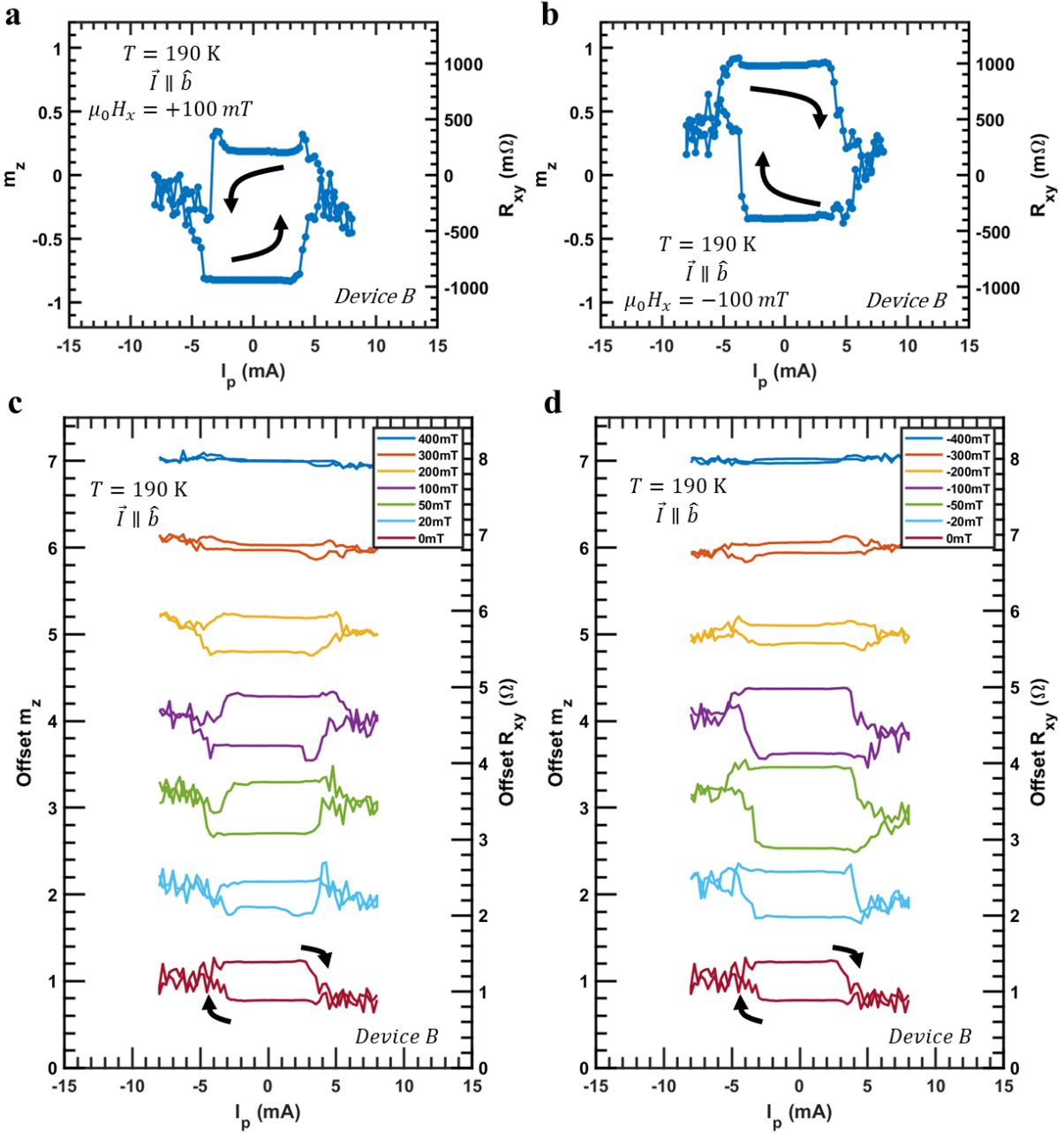

**Figure S14 | The chirality reversal of pulse-induced SOT switching of device B at various fields $H_x$ applied along the current. a,b,** The pulse-induced SOT switching at 190 K while $\vec{I}\|\hat{b}$ with $\mu_0 H_x$= +100 mT (**a**), and -100 mT (**b**) applied collinear to current direction. The chirality reverses when $H_x$ reverses. **c,d,** The pulse-induced SOT switching at 190 K while $\vec{I}\|\hat{b}$ at various fields with $H_x \geq 0$ (**c**), and $H_x \leq 0$ (**b**) applied collinear to current direction. When $|H_x| > 0$, the chirality is determined by the in-plane field, while the chirality is clockwise when $\mu_0 H_x = 0$ mT.

## Note 8. Current-Induced Field-Free Switching in Device C

Although the crystallographic orientation in device C is not verified by cross-sectional STEM or polarized Raman spectroscopy, the current channel is likely aligned to the a-axis of $WTe_2$ based on the cleaving directions of the flake and magnetotransport measurements. As shown in Figure S15a (b), similar to device A, we also observe field-free deterministic switching in device C with $m_z$ initially prepared in the negative (positive) state. The offset of the loops is likely due to an insufficient torque magnitude to bring the magnetization state to saturation.

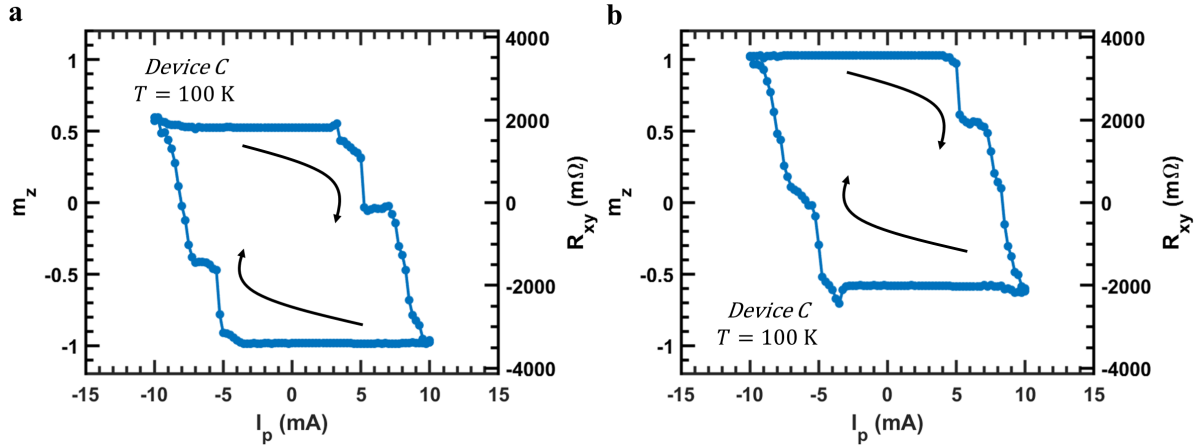

**Figure S15** | **The current-induced zero field SOT switching of device C. a,b,** The current-induced SOT switching at 100 K with $m_z$ initialized in negative **(a)** and positive **(b)** saturated state.

## Note 9. SOT Switching at Various Temperatures in WTe$_2$/FGT Heterostructures

We further demonstrate the deterministic SOT switching at various temperatures from 150 K to 190 K when $\vec{I} \parallel \hat{a}$ (Fig. S16a) and determine $J_{th}$, defined as $(J_{th}^+ + J_{th}^-)/2$ where $J_{th}^{+(-)}$ is the current density pulse magnitude that drives the magnetization state across $m_z = 0$ at positive (negative) pulse sides. The magnetization could be almost fully reversed from one saturated state to the opposite magnetization with $|m_z| \geq 0.88$. The non-saturated component is likely due to multi-domain formation due to Joule heating and the torques contributed by the Oersted effect and SHE that tend to drive the magnetization toward the plane. To compare the efficiency with similar vdW heterostructures, we plot the temperature dependence of the current density threshold $J_{th}$ from 150 K to 190 K (Fig. S16b). At lower temperatures, higher current is required to switch the magnetization due to the increase in $H_c$.

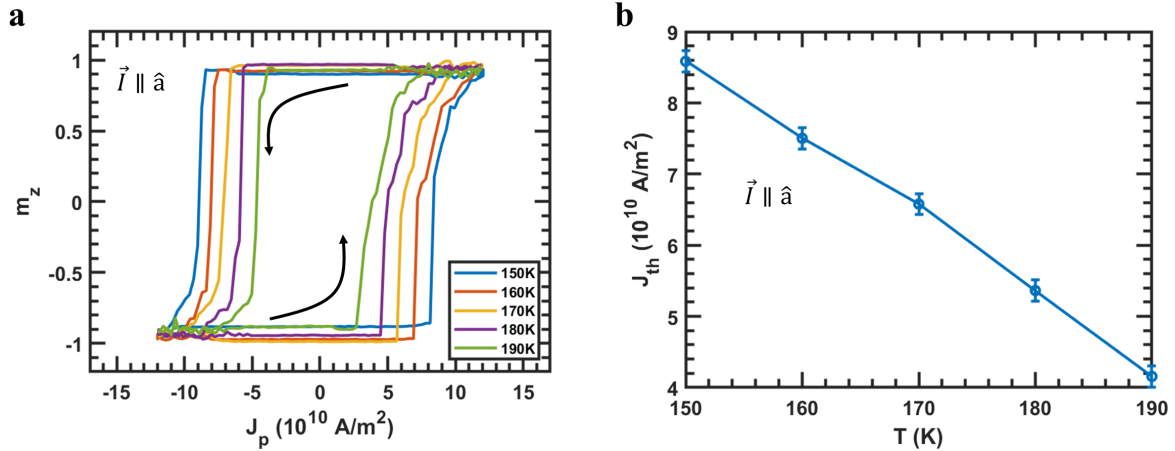

**Figure S16 | Current pulse-induced SOT switching and current density threshold at various temperatures of device A. a,** Current pulse-induced SOT switching full hysteresis loop while $\vec{I} \parallel \hat{a}$ at various temperatures from 150 K to 190 K. **b,** The current density threshold $J_{th}$ as a function of temperature.

# References

[S1] Song, Q. *et al.* The In-Plane Anisotropy of WTe$_2$ Investigated by Angle-Dependent and Polarized Raman Spectroscopy. *Scientific Reports* **6,** 29254 (2016).

[S2] Alghamdi, M. *et al.* Highly Efficient Spin–Orbit Torque and Switching of Layered Ferromagnet Fe$_3$GeTe$_2$. *Nano Letters* **19,** 4400–4405 (2019).

[S3] Fei, Z. *et al.* Two-dimensional itinerant ferromagnetism in atomically thin Fe$_3$GeTe$_2$. *Nature Materials* **17,** 778–782 (2018).

[S4] Wang, X. *et al.* Current-driven magnetization switching in a van der Waals ferromagnet Fe$_3$GeTe$_2$. *Science Advances* **5** (2019).

[S5] Li, P. *et al.* Evidence for topological type-II Weyl semimetal WTe$_2$. *Nature Communications* **8,** 2150 (2017).

[S6] Yi, J. *et al.* Competing antiferromagnetism in a quasi-2D itinerant ferromagnet: Fe$_3$GeTe$_2$. *2D Materials* **4,** 011005 (2016).

[S7] Zheng, G. *et al.* Gate-Tuned Interlayer Coupling in van der Waals Ferromagnet Fe$_3$GeTe$_2$ Nanoflakes. *Phys. Rev. Lett.* **125,** 047202 (2020).

[S8] Shao, Y. *et al.* The current modulation of anomalous Hall effect in van der Waals Fe$_3$GeTe$_2$/WTe$_2$ heterostructures. *Applied Physics Letters* **116,** 092401 (2020).

[S9] Liu, L., Lee, O. J., Gudmundsen, T. J., Ralph, D. C. & Buhrman, R. A. Current-Induced Switching of Perpendicularly Magnetized Magnetic Layers Using Spin Torque from the Spin Hall Effect. *Phys. Rev. Lett.* **109,** 096602 (2012).

[S10] Zhou, J., Qiao, J., Bournel, A. & Zhao, W. Intrinsic spin Hall conductivity of the semimetals MoTe$_2$ and WTe$_2$. *Phys. Rev. B* **99,** 060408 (2019).

[S11] MacNeill, D. *et al.* Control of spin-orbit torques through crystal symmetry in WTe$_2$/ferromagnet bilayers. *Nature Physics* **13,** 300–305 (2017).

[S12] MacNeill, D. *et al.* Thickness dependence of spin-orbit torques generated by WTe$_2$. *Phys Rev B* **96,** 054450 (2017).